\documentclass[sigconf,9pt]{acmart}
\pdfoutput=1
\usepackage{amsmath,amsfonts}
\usepackage[algo2e]{algorithm2e}
\usepackage{algorithm}
\usepackage{algcompatible}
\usepackage{tabularx}
\usepackage{subfig}
\usepackage{graphicx}
\usepackage{physics}
\usepackage{textcomp}
\usepackage{xcolor}
\usepackage{fancyhdr}
\usepackage{enumitem}
\renewcommand\footnotetextcopyrightpermission[1]{}

\newcommand{\sol}{\textsc{DisQ}}
\newcommand{\solc}{\textsc{C-DisQ}}
\newcommand{\sole}{\textsc{E-DisQ}}

\font\myfont=cmr12 at 12pt

\author{Tirthak Patel}
\affiliation{\institution{\myfont Northeastern University}}

\author{Devesh Tiwari}
\affiliation{\institution{\myfont Northeastern University}}

\begin{document}


\title{\sol{}: A Novel Quantum Output State Classification Method on IBM Quantum Computers using OpenPulse}


\begin{abstract}

Superconducting quantum computing technology has ushered in a new era of computational possibilities. While a considerable research effort has been geared toward improving the quantum technology and building the software stack to efficiently execute quantum algorithms with reduced error rate, effort toward optimizing how quantum output states are defined and classified for the purpose of reducing the error rate is still limited. To this end, this paper proposes \sol{}, a quantum output state classification approach which reduces error rates of quantum programs on NISQ devices.

\end{abstract}

\maketitle

\section{Introduction}
\label{sec:intro}

Quantum computing is advancing at a rapid pace with the proliferation of different quantum computing technologies and renewed interest from industry and academia. In addition to the better established quantum annealing approach which has limited applicability~\cite{pakin2016quantum,pakin2018survey,hassan2019c}, different quantum computing technologies are being actively explored to build a reliable quantum bit (or qubit), including superconducting qubits, trapped-ion qubits, and photon-based qubits. The most promising of these is the superconducting qubit technology which is primarily pioneered by IBM and Google. In fact, Google recently used their 53-qubit Sycamore quantum computing chip to run within a few seconds a task which would take a few days even on large supercomputers~\cite{arute2019quantum}.

State-of-the-art Noisy Intermediate-Scale Quantum (NISQ) computers based on superconducting quantum technology are actively being used to establish the usefulness of quantum computers with potential applications ranging from chemistry and physics simulations to hard optimization problems~\cite{bravyi2016trading,preskill2018quantum,martonosi2019next,huang2019statistical}. Unfortunately, due to the primitiveness of the technology, current NISQ computers have high error rates. Further, they do not have enough number of qubits to deploy meaningful error correction. Because of the high error rate of NISQ machines, the outputs generated by quantum algorithms run on current NISQ computers are erroneous. 

A significant amount of research effort has been geared toward understanding, debugging, and mitigating the errors of these computers. These works have mainly focused on three primary research aspects: (1) developing quantum error-and-noise-aware simulation frameworks~\cite{cirac2012goals,aleksandrowiczqiskit,mckay2018qiskit}, (2) debugging quantum programs to identify code-level errors~\cite{murali2019noise,liu2020quantum}, and (3) providing a best-effort solution to the NP-hard problem of mapping a logical quantum algorithm to a physical set of qubits such that the error rate of the produced output is minimized~\cite{bhattacharjee2019muqut,shi2019optimized,wille2019mapping,ash2019qure,li2019tackling,li2020towards,gokhale2019partial,tannu2019not,tannu2019ensemble,tannu2019mitigating,smith2019quantum,zulehner2019compiling,murali2020software}.

However, research efforts to address the problem of quantum output state classification are still in a nascent stage. This problem refers to the task of determining whether a qubit's output energy signal, after a quantum algorithm's execution, should be classified as the $\ket{0}$ state or the $\ket{1}$ state. This is not a straightforward task as the output state is affected by a variety of error-inducing factors.


\textit{To the best of our knowledge, we propose \sol{}, the first method to optimize the classifier which differentiates between different quantum output states using the quantum pulse schedules on an IBM superconducting quantum computer.}  \sol{} is built on two key insights: (1) The classification methodology, including  choice of quantum gates and classifier shape, can affect the error rate of quantum output; (2) More surprisingly, the quantum output error rate is dependent on the probability of the output states. Current state-of-art classifier is not aware of these characteristics and hence, suffers from a relatively high error rate. \sol{} addresses this challenge by building a classifier that is trained using multiple micro-benchmarks with different output state probabilities and optimized using a simulated annealing approach. We proposes two versions of \sol{}: circle- and ellipse-shaped classifiers. Our evaluation shows that \sol{} improves the median error rate, the 75$^{th}$ percentile error rate and the variability in error rate. \sol{} is practical and suitable for deployment: the training data can be collected within 2 minutes during calibration (calibration is the task of determining the optimal parameters to drive the qubit) and the classifier can be optimized within 7 minutes on an Intel Core i7 processor, and can be used for all algorithm runs until the next calibration period (approx. every 12 hours).

\textit{\sol{}'s classification toolbox is compatible with IBM's Python-based Qiskit OpenPulse framework and can be used with any IBM quantum computer which supports OpenPulse. It is available at: \texttt{http://github.com/GoodwillComputingLab/DISQ}.}
\section{Quantum Computing Background}
\label{sec:backg}

In this section, we provide some background of quantum computing, superconducting quantum computers, and their sources of errors.

\begin{figure}[t]
    \centering
    \subfloat[][]{\includegraphics[scale=0.22]{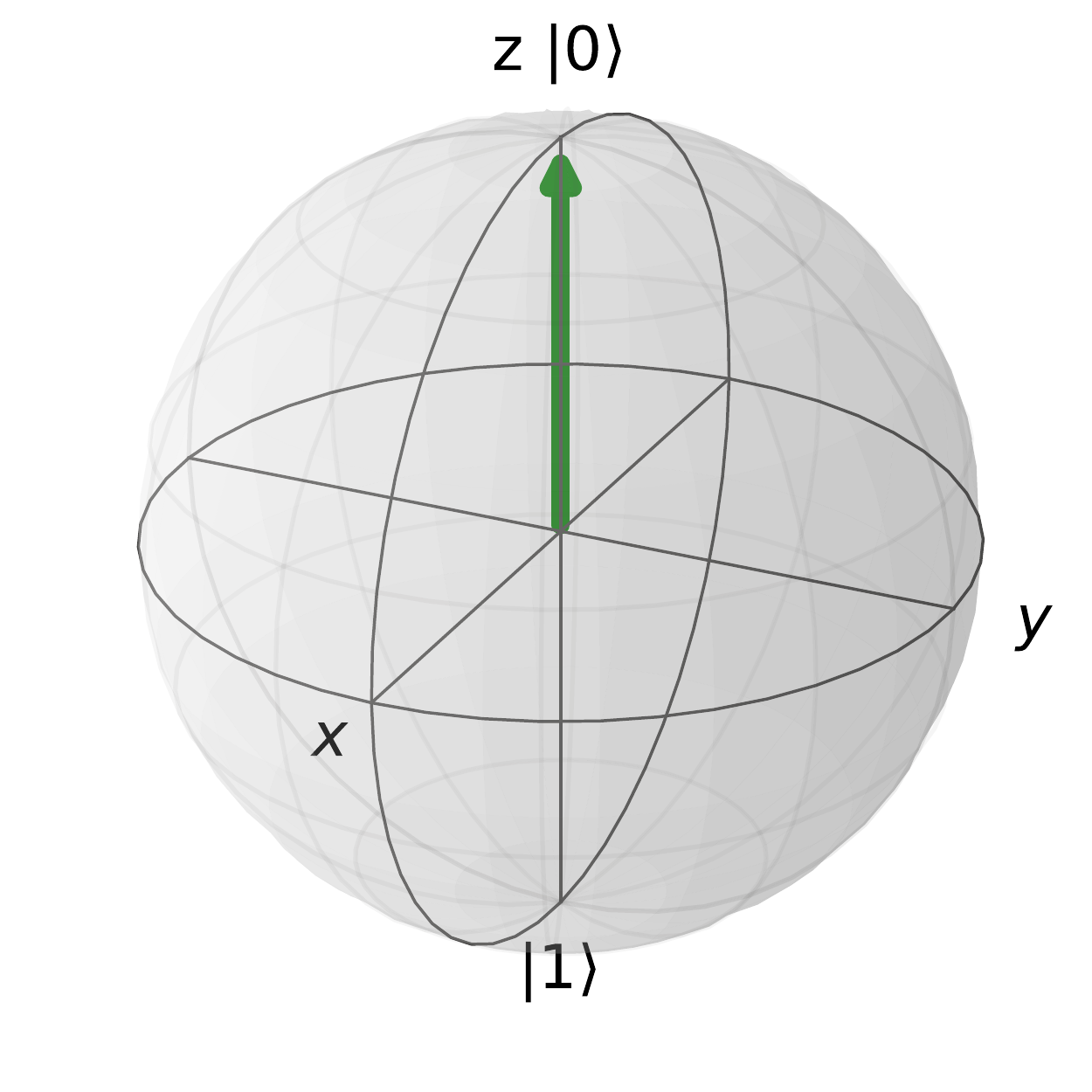}}
    \subfloat[][]{\includegraphics[scale=0.22]{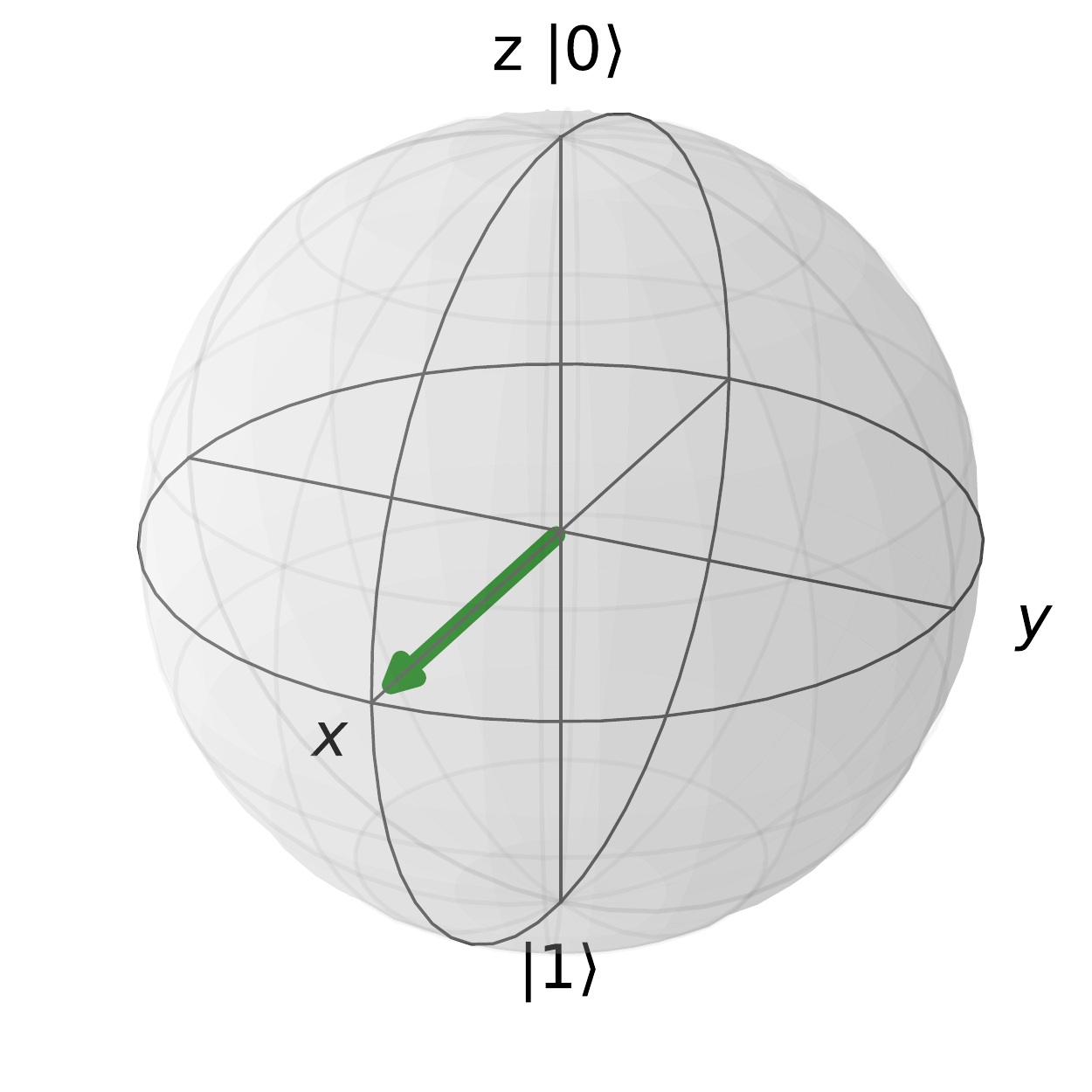}}
    \subfloat[][]{\includegraphics[scale=0.22]{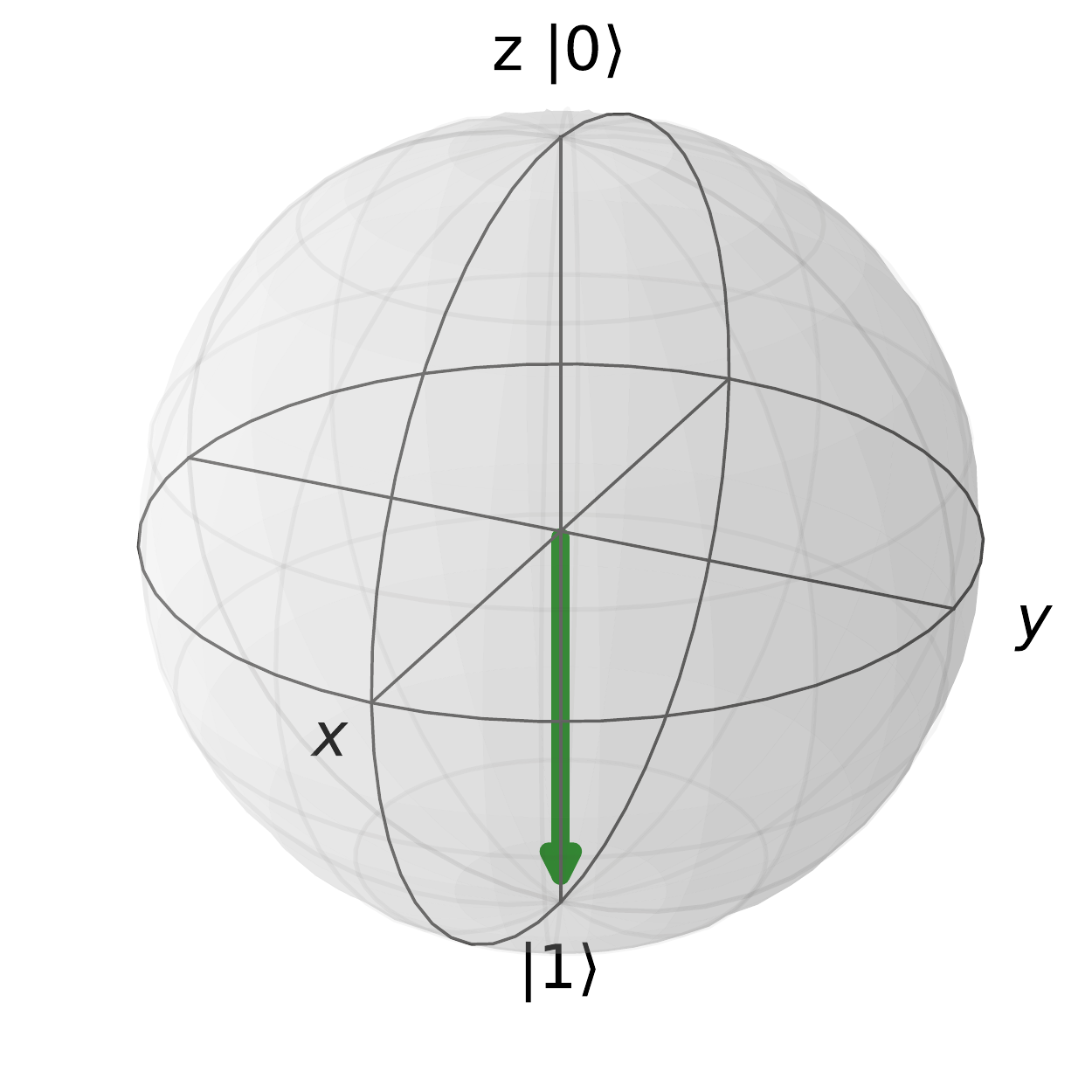}}
    \vspace{0.1cm}
    \hrule
    \vspace{-0.3cm}
    \caption{The Block Sphere visualizes a qubit's state. Here, two $R_y(\pi/2)$  gates are applied to a qubit initialized to state $\ket{0}$.}
    \vspace{-0.4cm}
    \label{fig:bloch}
\end{figure}

\begin{figure}[t]
    \centering
    \includegraphics[scale=0.32]{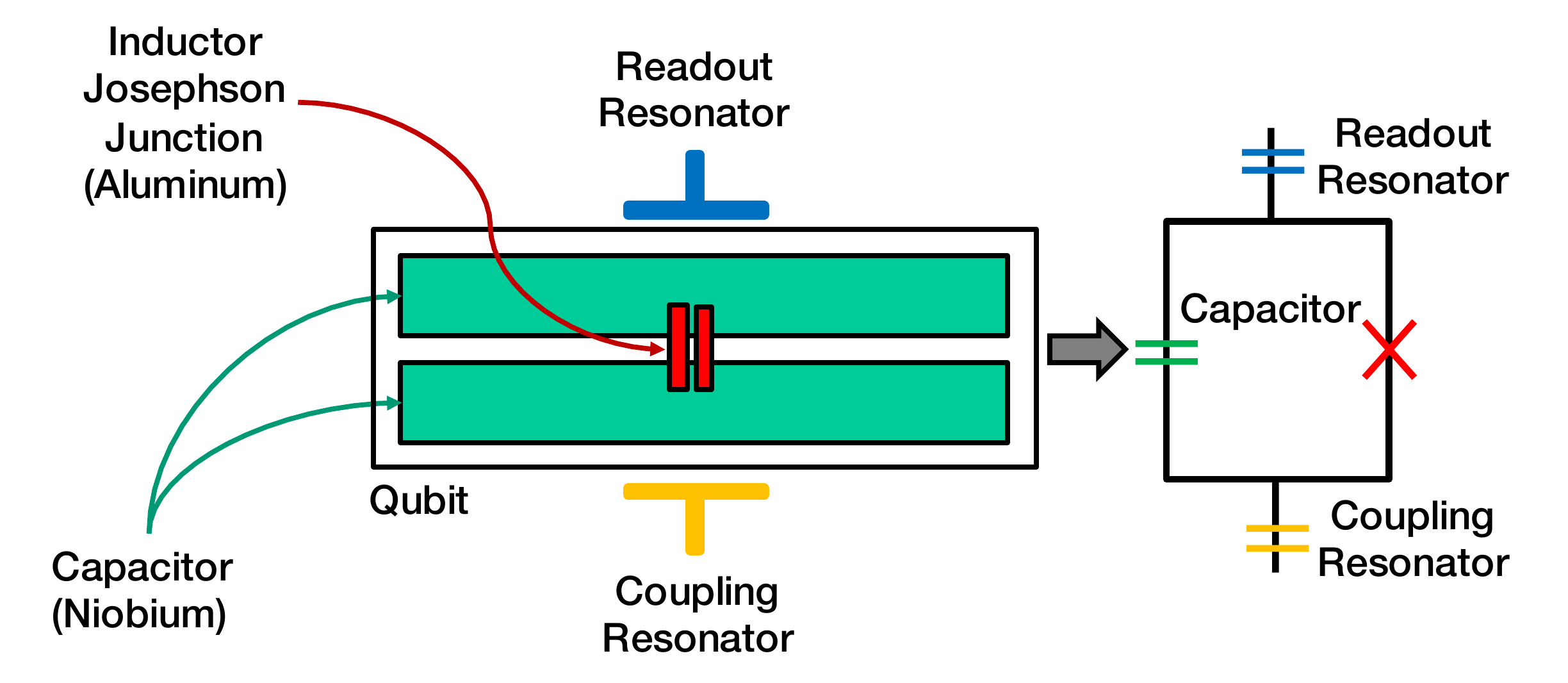}
    \vspace{0.1cm}
    \hrule
    \vspace{-0.3cm}
    \caption{A superconducting qubit is made out of an LC oscillator circuit constructed using a capacitor and a Josephson Junction.}
    \vspace{-0.4cm}
    \label{fig:super}
\end{figure}

\subsection{Quantum Computing Fundamentals}

 A \textit{qubit} is the building block of a quantum computer. A classical bit can exist in one of two states: $0$ or $1$. On the other hand, a \textit{qubit's state} ($\ket{\Psi}$), during computation, can be expressed as a \textit{superposition} of the two basis states: $\ket{0}$ and $\ket{1}$ (in bra-ket notation). More formally, $\ket{\Psi} = \alpha\ket{0} + \beta\ket{1}$, where $\alpha$ and $\beta$ are complex numbers such that  $\norm{\alpha}^2 + \norm{\beta}^2 = 1$. When the quantum computation is completed, the qubit's state is measured or \textit{read out}. When the qubit is read out, its superposition is destroyed, and it is measured in state $\ket{0}$ (with probability $\norm{\alpha}^2$) or in state $\ket{1}$ (with probability $\norm{\beta}^2$).
 
A \textit{quantum algorithm} is a set of \textit{quantum gates} sequentially applied to the qubits on a quantum computer. When a quantum algorithm completes execution, the state of all or some of the qubits is measured to analyze the output. A qubit's state can be read out only once during the algorithm execution as it collapses the qubit's superposition. Therefore, it is only read out at the end of the algorithm's execution.  Note that the output of a quantum algorithm is probabilistic. Therefore, multiple \textit{trials} are conducted to get the \textit{output probabilities} of a qubit state. For example, if a qubit has state $\ket{\Psi} = \frac{1}{\sqrt{2}}\ket{0} + \frac{1}{\sqrt{2}}\beta\ket{1}$, then the output probability of state $\ket{0}$ is $\norm{\frac{1}{\sqrt{2}}}^2 = \frac{1}{2}$ and state $\ket{1}$ is also $\norm{\frac{1}{\sqrt{2}}}^2 = \frac{1}{2}$. Therefore, if 1024 trials are conducted, then it would be expected that half of the trials (512) would have state $\ket{0}$ and the other half would have state $\ket{1}$.

Quantum gates can be of 1-qubit or 2-qubit variety. A 1-qubit gate operates on a single qubit and puts it in the desired superposition. A 2-qubit gate \textit{entangles} two qubits and modifies the superposition of the \textit{target} qubit based on the superposition of the \textit{control} qubit. A general 1-qubit gate has three components: x-, y-, and z- rotation components which rotate the qubit about the x-axis, y-axis and z-axis, respectively, on the Bloch Sphere (denoted as ($R_x(\theta)$, $R_y(\phi)$, and $R_z(\delta)$, respectively, where $\theta$, $\phi$, and $\delta$ are the angles of rotation). The Bloch Sphere is a unit sphere with the $\ket{0}$ state represented as a vector pointing toward the positive z-axis and the $\ket{1}$ state represented on the negative z-axis. While, the qubit state vector can be pointed in any direction on the Bloch Sphere, when it is measured, it collapses to the $\ket{0}$ or the $\ket{1}$ state, and its output probability is the projection of the vector onto the z-axis. For example, in Fig.~\ref{fig:bloch}, the Bloch Sphere shows the state changes after applying gates to a qubit initialized to state $\ket{\psi} = \ket{0}$. First, a $R_y(\pi/2)$ gate is applied. This puts the qubit in equal superposition ($0.5$ probability of observing both $\ket{0}$ and $\ket{1}$). Another $R_y(\pi/2)$ gate brings the qubit to state $\ket{1}$. Arbitrary rotations can be applied about any of the three axes to achieve the desired superposition. All 1-qubit rotations have 2-qubit variants ($CR_x(\theta)$, $CR_y(\phi)$ and $CR_z(\delta)$), where the target qubit is rotated as per the control qubit.

\subsection{IBM's Superconducting Qubit Technology}

IBM uses a circuit-based approach toward developing qubits using superconducting Josephson Junctions. As shown in Fig.~\ref{fig:super}, a superconducting niobium linear capacitor and a superconducting aluminium Josephson Junction, which behaves as a non-linear non-dissipative inductor, are developed on a silicon wafer. To build the Josephson Junction, two pieces of weakly coupled superconducting electrodes are separated by a very thin tunnel barrier which serves as an insulating layer. Overall, this forms a non-linear LC oscillator which behaves like a qubit if the parameters are tuned correctly. The oscillator allows for a two-level quantum system with discrete quantum energy levels. The lowest two energy levels are used as the $\ket{0}$ (ground level) and the $\ket{1}$ (first excited level) states.

\begin{figure}[t]
    \centering
     \includegraphics[scale=0.25]{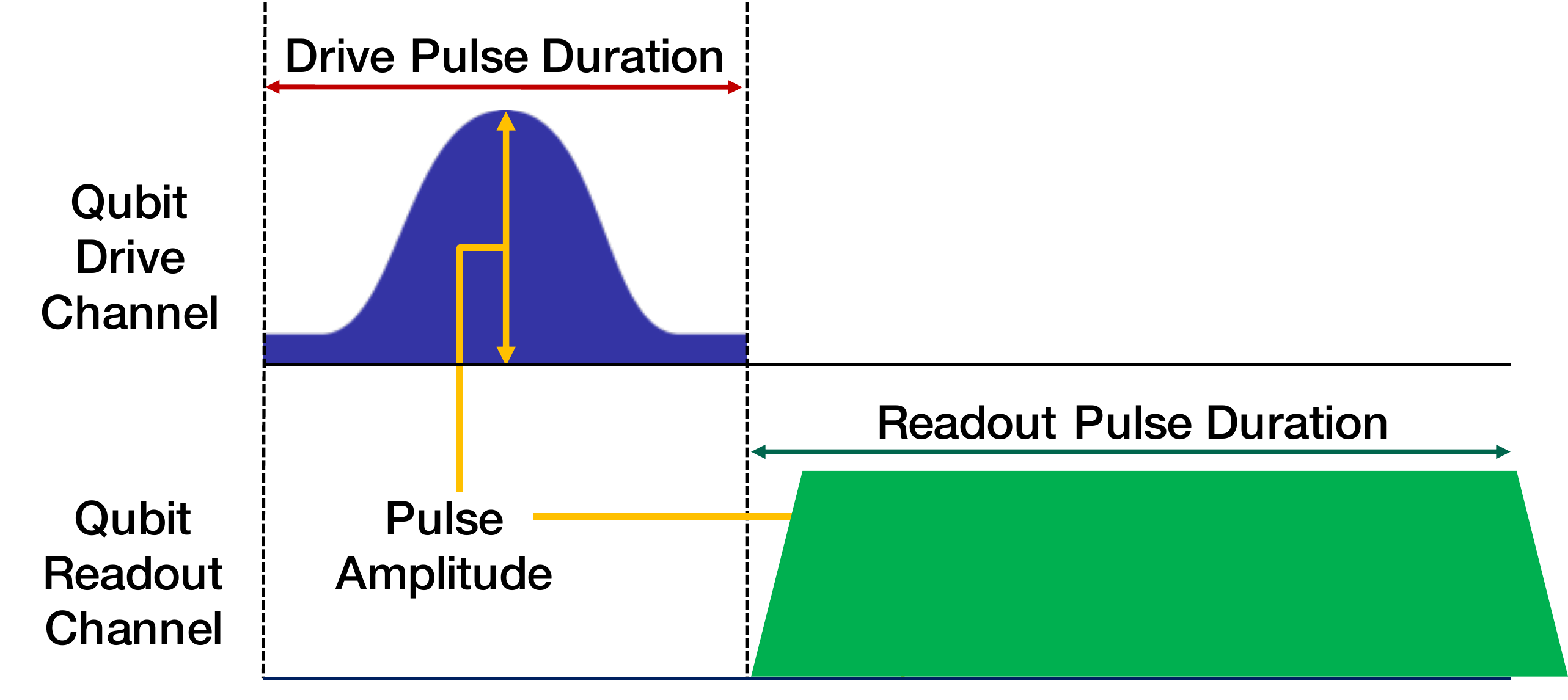}
    \vspace{0.1cm}
    \hrule
    \vspace{-0.3cm}
    \caption{Microwave pulses are applied to drive the qubit to a desired state and read out the qubit value.}
    \vspace{-0.4cm}
    \label{fig:pulse}
\end{figure}

\subsection{Driving and Measuring Qubit States}

The superconducting-qubit system can be addressed using external controls. Typically, the superconducting qubit has a frequency in the range of 4-5 GHz. As shown in Fig.~\ref{fig:pulse}, a Microwave tone can be applied at the qubit frequency to drive the qubit. By applying a pulse shape to the Microwave tone, quantum gates can be achieved. Typically, a Microwave pulse with a Gaussian shape is used to drive the qubit. Frame-of-reference change is used to apply gates along different axes. The pulse with the maximum amplitude magnitude is known as the ``$\pi$ Pulse'' and it is used to apply the $R_x(\pi)$ rotation, which transforms the $\ket{0}$ state to the $\ket{1}$ state and vice versa.

The qubit state is measured by coupling the qubit to a superconducting transmission resonator. It is ensured that the readout resonator frequency is dependent on the state of the qubit. Thus, the qubit state can be determined by probing the resonant frequency. Fig.~\ref{fig:pulse} shows that a long Square Gaussian pulse is applied to measure the state of the qubit. Lastly, as shown in Fig.~\ref{fig:super}, 2-qubit entangling gates are applied using a superconducting coupling bus resonator.

\subsection{Sources of Error in Quantum Computers}

There are several sources of error in NISQ technology. Once initialized, a qubit can only hold the coherence of its state for a limited amount of time. There are two types of coherence decays: (1) The T1 coherence refers to the amplitude damping. (2) The T2 coherence refers to the phase damping. NISQ errors also include the gate and readout errors. An erroneous application of the microwave pulses can cause gate errors, i.e., the gate could be put in a slightly-off-the-desired superposition which can lead to incorrect output probabilities. The readout resonators are also error-prone and cause readout errors. Refer to Patel et al.~\cite{patel2020ureqa} for more details about the different sources of error. These errors can be a result of deviations from the optimal pulse parameters such as the frequency at which the pulse is applied, its amplitude, and its duration. Moreover, because the technology is still maturing, the qubit's properties, such as its frequency, vary. Therefore, the optimal pulse parameters need to be determined on a regular basis.

For this reason, IBM's computers are calibrated twice a day, and the qubit's error rates change after each calibration. \textit{Calibration} is the task of determining qubit properties, such as frequency, and accordingly, setting the parameters of the microwave pulses for the gate operations. These parameters are then used to perform all the operations until the next calibration. Also, during the calibration period, the classifier which distinguishes between the $\ket{0}$ and the $\ket{1}$ state is developed (as discussed next in Sec.~\ref{sec:motiv}).
\section{Motivation for \sol{}}
\label{sec:motiv}

In this section, we introduce the state classification problem, the existing state-of-art classification method, and its inefficiencies.

\begin{figure}[t]
    \centering
    \subfloat[][Hadamard Output]{\includegraphics[scale=0.435]{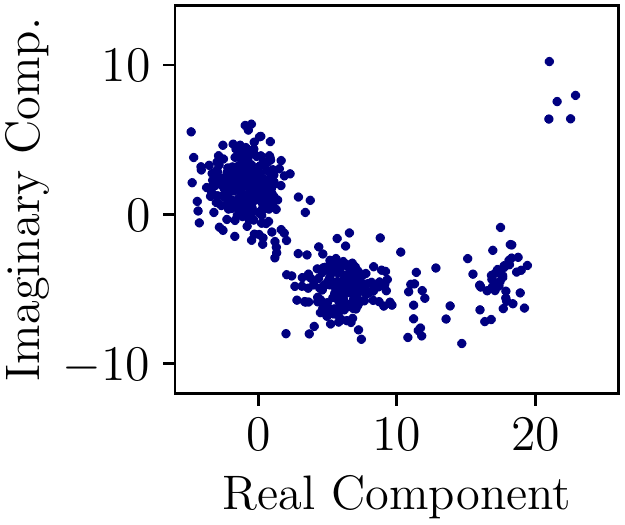}}
    \hfill
    \subfloat[][Good Discriminant]{\includegraphics[scale=0.435]{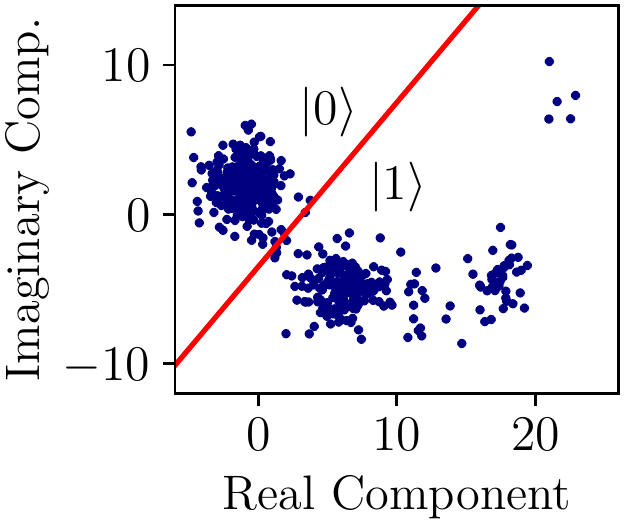}}
    \hfill
    \subfloat[][Poor Discriminant]{\includegraphics[scale=0.435]{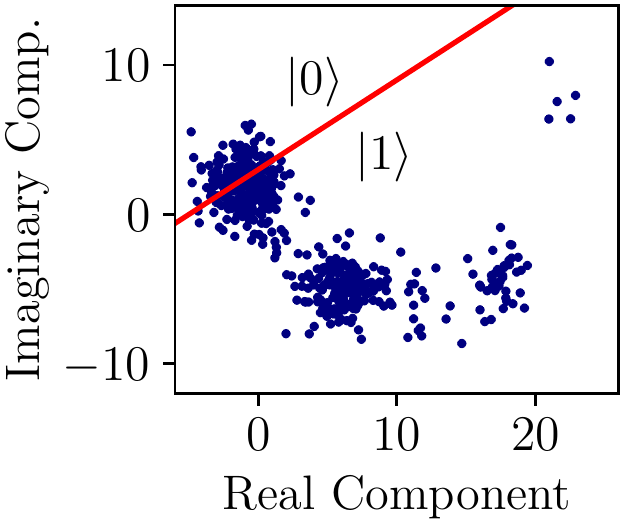}}
    \vspace{0.1cm}
    \hrule
    \vspace{-0.3cm}
    \caption{The role of a classifier is to identify the outputs of trials as $\ket{0}$ or $\ket{1}$ in a manner which minimizes the error.}
    \vspace{-0.4cm}
    \label{fig:cladef}
\end{figure}

\begin{figure}[t]
    \centering
    \subfloat[][Step 1: State $\ket{0}$ Run]{\includegraphics[scale=0.435]{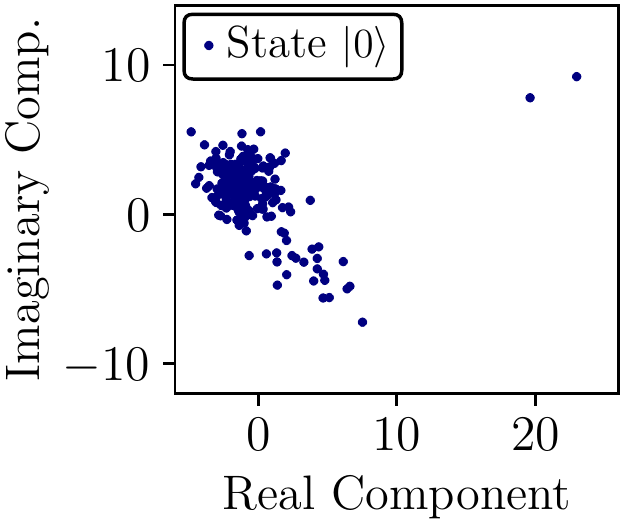}}
    \hfill
    \subfloat[][Step 2: State $\ket{1}$ Run]{\includegraphics[scale=0.435]{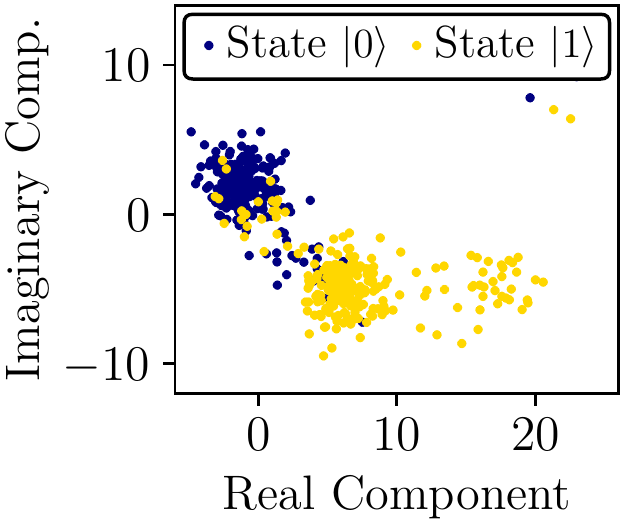}}
    \hfill
    \subfloat[][Step 3: Discriminant]{\includegraphics[scale=0.435]{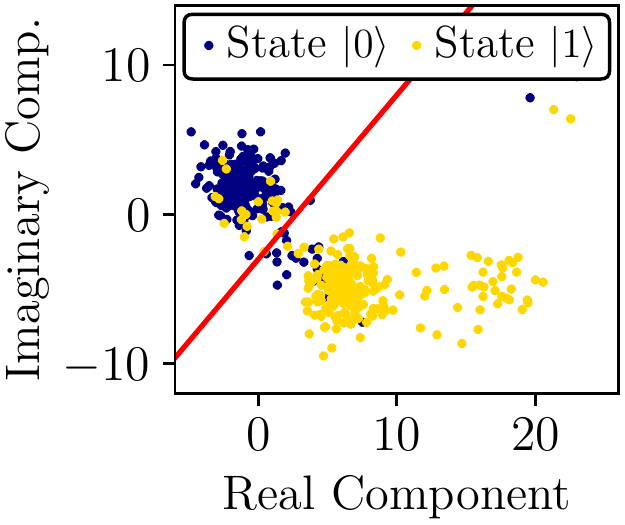}}
    \vspace{0.1cm}
    \hrule
    \vspace{-0.3cm}
    \caption{In the baseline case, a linear discriminant is used to differentiate the $\ket{0}$ vs. $\ket{1}$ using a three-step process.}
    \vspace{-0.4cm}
    \label{fig:lincla}
\end{figure}

\subsection{Classification of Output States}

Once a quantum algorithm has completed execution, its final state is measured. To do this, the readout pulse is run on the readout channel (as shown in Fig.~\ref{fig:pulse}), and the corresponding signal, which measures the energy state of the qubit, is recorded on the acquire channel. The signal recorded during the entire acquire duration is then summed up and a single value is returned for each trial for each qubit. This value is a complex number with a real and an imaginary component. It is used to determine if the measured state represents the $\ket{0}$ state or the $\ket{1}$ state.

As an example, consider that a quantum gate known as the Hadamard gate is executed on a qubit initialized to state $\ket{0}$ and its resulting final state is read out. The resulting output values of conducting 1024 trials are shown in Fig.~\ref{fig:cladef}(a). The Hadamard gate puts a qubit in equal superposition of the $\ket{0}$ and the $\ket{1}$ state (0.5 probability of both states). Therefore, ideally, 512 of these trial outputs should be classified as state $\ket{0}$ and the other 512 as state $\ket{1}$. Given the nature of NISQ computers, the ideal classification is not possible. However, best effort must be made to reduce the \textit{output error}. The output error (in percentage) is defined as the |\textit{correct probability of state $\ket{0} -$ observed probability of state $\ket{0}$}|$\times$100 = |\textit{correct probability of state $\ket{1} -$ observed probability of state $\ket{1}$}|$\times$100. The observed probability of $\ket{0}$ is calculated by dividing the number of trials which resulted in $\ket{0}$ by the total number of trials conducted. Also, by definition, probability of $\ket{1} = 1 -$ probability of $\ket{0}$.

A trivial choice for state classification is a linear classifier. Consider the linear classifier used in Fig.~\ref{fig:cladef}(b) which classifies 548 trial outputs as state $\ket{0}$. Then, the observed output probability of state $\ket{0}$ is $548/1024 = 0.535$, which results in an output error of $|0.5 - 0.535|\times100 = 3.5\%$. A poorer quality classifier shown in Fig.~\ref{fig:cladef}(c) classifies 323 trial outputs as state $\ket{0}$, resulting in an error of 18.5\%. Thus, developing a classifier which minimizes the output error is critical. Note that it is not possible to use different classifiers which cater to the output probabilities of different quantum algorithms as the output probabilities are not known for real quantum algorithms. Therefore, only one classifier is developed and used for all algorithms. Typically, the classifier is developed during calibration, once the qubit's frequency and pulse parameters, such as amplitude, are determined.

\subsection{IBM's Existing Classification Methodology}

Next, we look at the state-of-art method used for classification~\cite{aleksandrowiczqiskit}.

\textbf{Step I.} First, the qubit is initialized to the $\ket{0}$ state and its value is directly readout without running any gate operations. This is done for multiple trials. The probability of observing state $\ket{0}$ would be 1 in this case. Therefore, ideally, the output value of all trials should be classified as belonging to state $\ket{0}$. Fig.~\ref{fig:lincla}(a) shows the output spattering of 1024 trials of this run on the complex plane.

\textbf{Step II.} Next, the output values for the $\ket{1}$ state need to be generated. To achieve this, the qubit is initialized to the $\ket{0}$ state. Then, a $\pi$ pulse (i.e., a rotation of $\pi$ about the x-axis -- $R_x(\pi)$) is applied to it to put it in the $\ket{1}$ state (pointing toward the negative z-axis in the Bloch Sphere). It is then measured, as shown in Fig.~\ref{fig:pulse}. Ideally, the output value of all trials of this run should be classified as belonging to the $\ket{1}$ state. Fig.~\ref{fig:lincla}(b) shows the output spattering of 1024 trials of this run which produces the $\ket{1}$ state.

\textbf{Step III.} The last step is to construct a discriminator. Fig.~\ref{fig:lincla}(b) shows that there could be some overlap between the values generated by the $\ket{0}$ state and the values generated by the $\ket{1}$ state. This is due to the various gate and coherence errors mentioned in Sec.~\ref{sec:backg} which contribute to erroneous output. Therefore, a perfect distinction between the $\ket{0}$ and the $\ket{1}$ state, which would result in an output error of 0\%, is not possible to achieve. However, best effort must be made to reduce the output error. Currently, a linear discriminator is used to minimize this output error by generating a discriminator which maximizes the distance between the means of the classified samples, as shown in Fig.~\ref{fig:lincla}(c). Points that fall above the line are classified as $\ket{0}$ and ones that fall below the line are classified as $\ket{1}$. For all the runs conducted until the next calibration, this linear discriminator is used to classify the qubit state.

\begin{figure}[t]
    \centering
    \includegraphics[scale=0.47]{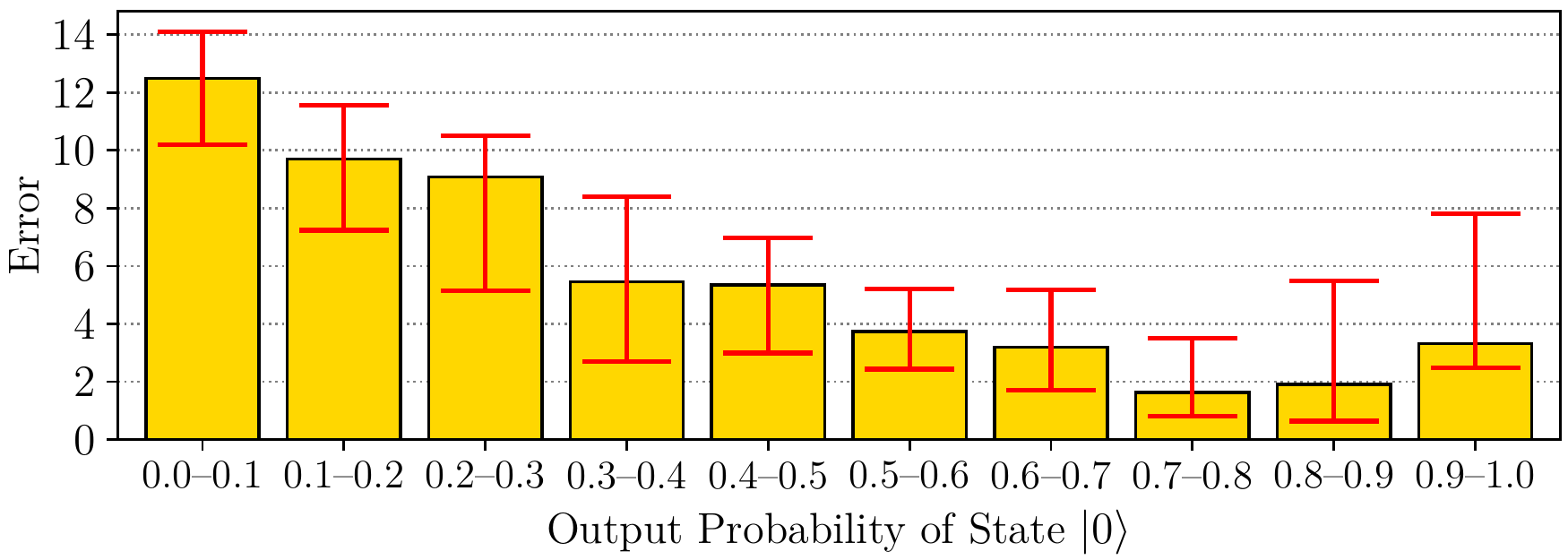}
    \vspace{0.1cm}
    \hrule
    \vspace{-0.3cm}
    \caption{Different error rates are observed for runs with different output probabilities of state $\ket{0}$. The bars show the median in each bin; the range indicators show the spread from the 25$^{th}$ percentile error to the 75$^{th}$ percentile error.}
    \vspace{-0.4cm}
    \label{fig:defbin}
\end{figure}

\begin{figure}[t]
    \centering
    \includegraphics[scale=0.3]{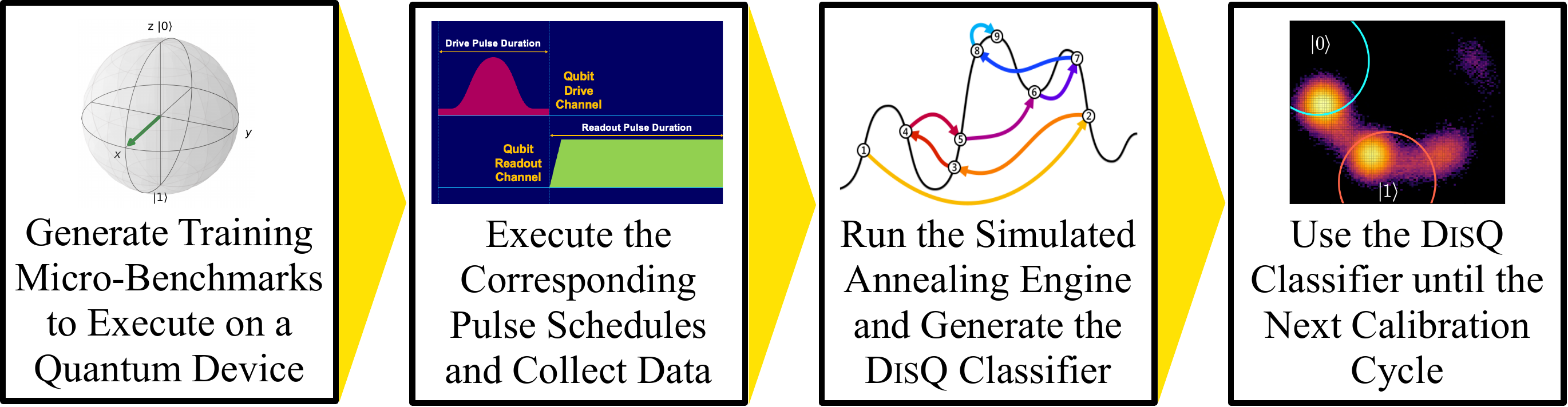}
    \vspace{0.1cm}
    \hrule
    \vspace{-0.3cm}
    \caption{Overview of the steps when classifying states using \sol{}.}
    \vspace{-0.4cm}
    \label{fig:discos}
\end{figure}

\subsection{Inefficiencies of the Existing Method}

The existing state classification method or the \textit{baseline} method uses only two runs -- one with output state $\ket{0}$ with probability 1 and the other with output state $\ket{1}$ with probability 1 -- to develop the linear discriminant. As such, applying this discriminant to runs with other output probabilities can lead to several undesired consequences. To demonstrate this, we ran over one thousand micro-benchmark runs for a period of 10 days on IBM's 1-qubit Armonk quantum computer. For each run, a $U3(R_x(\theta)$, $R_y(\phi)$, $R_z(\delta))$ gate with three random rotation angles about each of the three axes, selected uniformly between $-\pi$ and $\pi$, is applied to a qubit initialized to state $\ket{0}$. This results in an arbitrary output probability of state $\ket{0}$ between 0 and 1. Fig.~\ref{fig:defbin} shows the error rates of runs when they are binned in increments of 0.1 of their correct output probabilities. We ensured that all bins have the same number of runs. Fig.~\ref{fig:defbin} shows several interesting results. \newline

\noindent\textit{(1) The output error is highly correlated with the output probability of state $\ket{0}$.} For example, the median error when the output probability of state $\ket{0}$ is 0.9-1 is 3.5\% but the median error when the output probability of state $\ket{0}$ is 0-0.1 is 12.5\%, which is 3.5$\times$ worse compared to the former's error rate. Some of this difference is due to the fact that when the output probability of state $\ket{0}$ is 0-0.1, it means that state $\ket{1}$ is expected for most of the trials. But, due to the coherence properties of qubits, state $\ket{1}$ is more likely to lose its coherence and drop from the first excited state to the ground state. However, a degradation in error rate of 3.5$\times$ is unacceptable. Quantum algorithms with lower output probability of state $\ket{0}$ should not observe a much higher error rate simply due to the makeup of their output probabilities. An algorithm's output probability of states cannot be controlled or modified as it is an intrinsic property of the algorithm. Therefore, the output state classifier should be developed in accordance with these unintentional outcomes, such that similar error rates are observed regardless of the output probability of state $\ket{0}$.\newline

\noindent\textit{(2) The error rate is high even for a single U3 gate used for the micro-benchmarks; a typical quantum algorithm consists of many gate operations and these errors get compounded.} The median error across all the runs is about 5\%, while the $75^{th}$ percentile error is 9\%. While the error rate is high in isolation, note that the error rate compounds when more gates are applied as each gate operation introduces its own error rate while also increasing the probability of state losing coherence as more time is spent before measuring the qubit state. Therefore, effort is needed to decrease the median error rate and especially, the $75^{th}$ percentile error rate as 25\% of the runs have more than 9\% error rate.\newline 

\noindent\textit{(3) The spread or variability in error rate is high even among runs with similar output probabilities of state $\ket{0}$.} For example, in Fig.~\ref{fig:defbin}, consider the 0.2-0.3 bin. It has a median error of 8.75\%. But it has a significant error spread, i.e., the 75$^{th}$ percentile error - the 25$^{th}$ percentile error, of 6.5\%. Similarly, other bins also have a high spread. The fact that the error rate varies so much even among runs with similar output probability of state $\ket{0}$ indicates that the error rates are not stable and the results are non-reproducible. Therefore, it needs to be ensured the spread of the error rates is minimized.  

\section{\sol{} State Classification Design}
\label{sec:design}

In this section, we describe \sol{}, an approach to mitigate the aforementioned issues in Sec.~\ref{sec:motiv}. Fig.~\ref{fig:discos} provides a high-level overview of the procedure that \sol{} uses to develop a discriminating state classifier. When a qubit is being calibrated, \sol{} should be executed to obtain the state classifier. First, \sol{} generates different micro-benchmarks with a diverse range of output probabilities of state $\ket{0}$ to cover the entire spectrum of possible output state probabilities. Next, it executes these micro-benchmarks on the qubit which needs to be calibrated and obtains the raw complex data. Post this, \sol{} runs a black-box simulated annealing engine which optimizes the parameters of the discriminating classifier in a manner which minimizes the median error \textit{and} the spread of the error of the training data. This classifier can then be used for all quantum algorithms executed on that qubit until the next calibration is performed.

\begin{figure}
    \centering
    \includegraphics[scale=0.3]{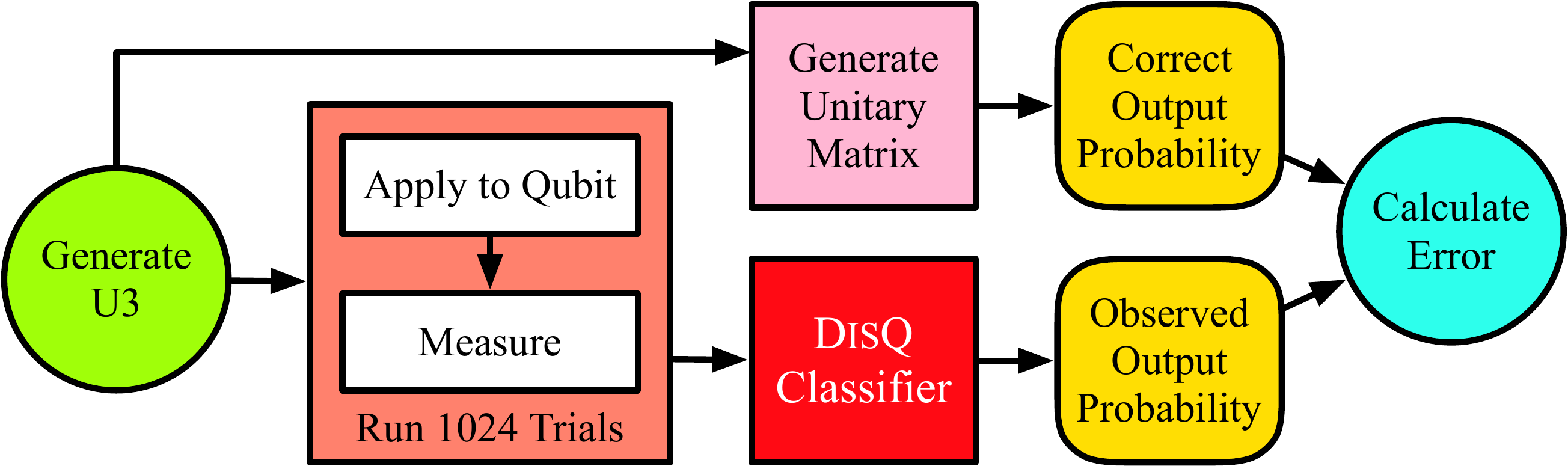}
    \vspace{0.1cm}
    \hrule
    \vspace{-0.3cm}
    \caption{Processing and calculating error of a micro-benchmark.}
    \vspace{-0.4cm}
    \label{fig:u3flow}
\end{figure}

\subsection{Micro-Benchmark Design and Execution}

The first order of business is to construct the micro-benchmarks whose output is fed to the state-classifier for discriminating the $\ket{0}$ and the $\ket{1}$ states. The purpose of these micro-benchmarks is to cover the full spectrum of output probabilities of state $\ket{0}$ from 0 to 1. As discussed earlier, the reason for this purpose is to ensure that the developed classifier minimizes and achieves equal error rate for quantum algorithms with all types of output probabilities. 

We achieve this on IBM's Qiskit framework by applying the $U3(R_x(\theta)$, $R_y(\phi)$, $R_z(\delta))$ gate with three random rotation angles about each of the three axes, selected uniformly between $-\pi$ and $\pi$, to a qubit initialized to state $\ket{0}$. This results in an arbitrary correct output probability of state $\ket{0}$ between 0 and 1 which can be calculated from the corresponding unitary matrix of the $U3$ gate. We run multiple such randomized micro-benchmarks. We ensure that each output probability of state $\ket{0}$ bin in increments of 0.1 (i.e., $0-0.1, 0.1-0.2,\dots,0.9-1$) contains equal number of micro-benchmarks. These micro-benchmarks are then executed on a quantum computer with 1024 trials each, their observed output probability is calculated using \sol{} classifier, and their output error is calculated, as shown in Fig.~\ref{fig:u3flow}. The next step for \sol{} is to classify.

\begin{figure}[t]
    \centering
    \subfloat[][Linear Classifier]{\includegraphics[scale=0.54]{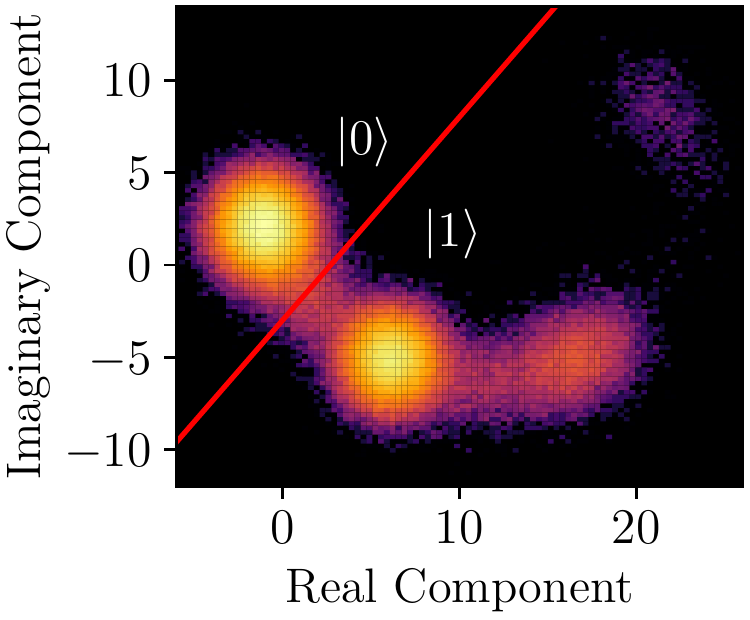}}
    \hfill
    \subfloat[][High-Overlap Zones]{\includegraphics[scale=0.173]{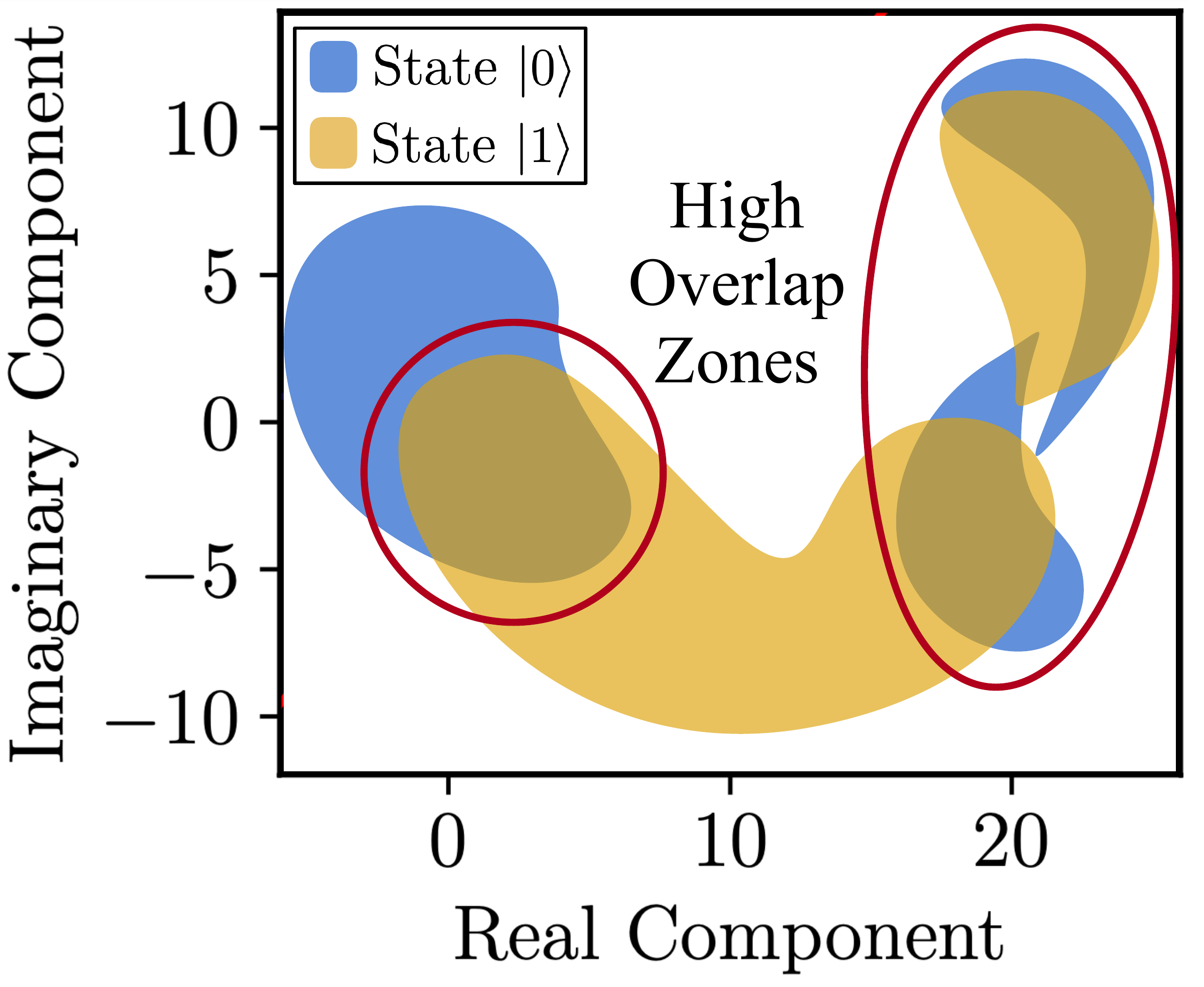}}
    \vspace{0.1cm}
    \hrule
    \vspace{-0.3cm}
    \caption{A linear classifier does not take into account high-overlap zones which are error-prone because they have a similar density of points belonging to the $\ket{0}$ and $\ket{1}$ states.}
    \vspace{-0.6cm}
    \label{fig:overlp}
\end{figure}

\begin{figure}[t]
    \centering
    \subfloat[][\solc{}: Circle Classifier]{\includegraphics[scale=0.54]{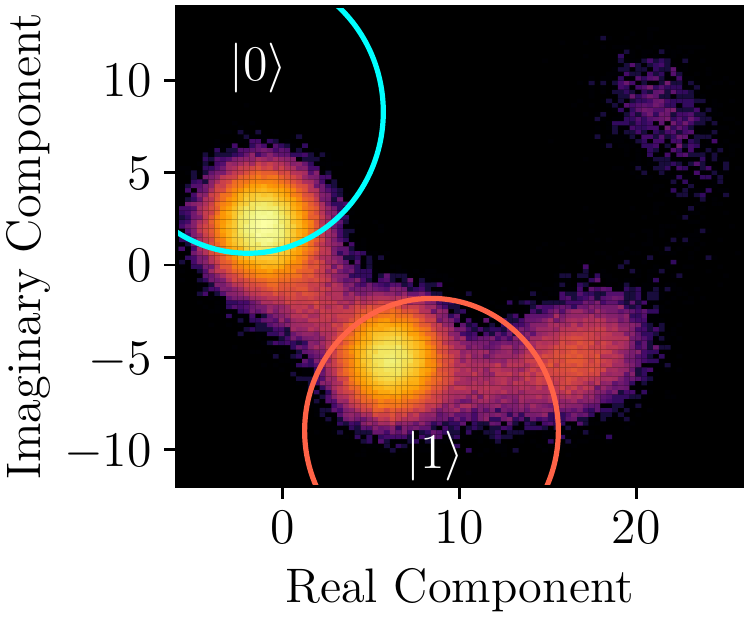}}
    \hfill
    \subfloat[][\sole{}: Ellipse Classifier]{\includegraphics[scale=0.54]{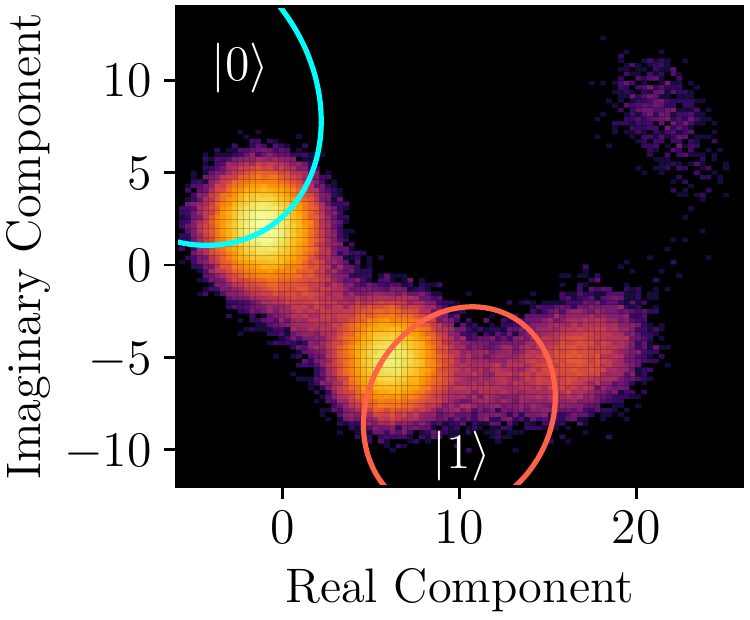}}
    \vspace{0.1cm}
    \hrule
    \vspace{-0.3cm}
    \caption{Visual representation of \solc{} (circle classifier) and \sole{} (ellipse classifier), which focus on low-overlap zones.}
    \vspace{-0.4cm}
    \label{fig:cirell}
\end{figure}

\begin{algorithm}[t]
{\fontsize{8pt}{8pt}\selectfont
\begin{algorithmic}[1]
\STATE \textbf{Input:} Training data $T_{\text{data}}$, number of iteration $N_{\text{iter}}$, initial temperature $T$, and cooling coefficient $\alpha$
\STATE Best configuration $C_{best}$ (sampled randomly)
\STATE Best objective $P_{\text{best}} \Leftarrow \text{Objective}(T_{\text{data}}, C_{\text{best}})$
\FOR {$i = 1,\dots,N_{\text{iters}}$}
\STATE Random neighbor configuration $C_{\text{new}}$. 
\STATE Corresponding $P_{\text{new}} \Leftarrow \text{Objective}(T_{\text{data}}, C_{\text{new}})$.
\STATE Energy $E \leftarrow P_{\text{best}} - P_{\text{new}}$.
\IF {$E > 0$ \textbf{or} $E < 0$ \textbf{and} $\text{random}() < e^{E / T}$}
\STATE $C_{\text{best}} \Leftarrow C_{\text{new}}$
\STATE $P_{\text{best}} \Leftarrow P_{\text{new}}$
\ENDIF
\STATE $T \Leftarrow \alpha \times T$
\ENDFOR
\STATE \textbf{Output} Best discriminator configuration $C_{\text{best}}$ 
\end{algorithmic}}
\caption{\sol{}'s simulated annealing engine.}
\label{alg:discsa}
\end{algorithm}

\subsection{\sol{}'s Discriminating Classifiers}

The primary consideration when developing a discriminating classifier is to determine the shape of the classifier. The existing approach is to use a linear discriminant as shown in Fig.~\ref{fig:overlp}(a), which shows an example density plot of the output of multiple training micro-benchmarks. The points above the line are classified as state $\ket{0}$, and the ones below are classified as state $\ket{1}$. Thus, a linear discriminant classifies points generated by all the trials that are conducted.

\textit{However, during the design of \sol{},  our experimental results revealed that it might not be suitable to use all the trial outputs}. As Fig.~\ref{fig:overlp}(b) shows, there are certain zones on the complex plane where both, points belonging to state $\ket{0}$ and points belonging to state $\ket{1}$, exist with an equal density. For example, 45\% of the points falling in these regions could represent state $\ket{0}$ and the remaining 65\% could represent state $\ket{1}$. Thus, these zones are not ideal for classification purposes as points in these regions are highly likely to be $\ket{0}$ or $\ket{1}$. \textit{Informed by these observations, we designed \sol{} to ignore such high-overlap zones and instead, focus on low-overlap zones.} For example, such low-overlap zones could be where point density of state $\ket{0}$ is 99\% and that of state $\ket{1}$ is 1\%.

A linear classifier cannot achieve this. In fact, any line-based classifier (quadratic, cubic, etc.) accounts for all trials by design. \textit{To solve this problem, we propose two classifiers which can identify low-overlap zones and not consider high-overlap zones: a circle discriminant classifier (\solc{}) and an ellipse discriminant classifier (\sole{}}). Examples of both of these classifiers are given in Fig.~\ref{fig:cirell}. As shown in the figure, both of these classifiers can focus on low-overlap zones and avoid high-overlap zones. For example, consider an algorithm that is run with 1024 trials, and 400 fall in the $\ket{0}$ region, 500 fall in the $\ket{1}$ region, and 124 fall outside both regions. Then, those 124 trials can be ignored, and instead the observed output probability of state $\ket{0}$ can be calculated as $400/(400+500) = 0.44$. This is in contrast to a linear classifier where all trails have to be considered, even ones which fall in high-overlap error-prone zones.

However, it is non-trivial to determine the parameters which define the \solc{} and \sole{} classifiers. \solc{} has six parameters, three for each circle (x-coordinate, y-coordinate, and radius). \sole{} has ten parameters, five for each of the ellipses (x-coordinate, y-coordinate, width, height, and angle of rotation). While, \solc{} has fewer parameters to optimize, \sole{} has better precision over the classification region which can potentially lead to a lower error rate (both are evaluated in Sec.~\ref{sec:eval}). Nonetheless, the problem of determining their parameters, while minimizing the error rate and error spread calculated based on the micro-benchmark measured data, is NP-hard. Therefore, next, we discuss the black-box simulated annealing engine used to optimize the two classifiers.

\subsection{\sol{} Simulated Annealing Engine}

The goal of \sol{}'s simulated annealing engine is to obtain the \textit{parameter configuration} for \solc{} and \sole{} such that the given objective is nearly-minimized (for the micro-benchmark measured data). A parameter configuration is one permutation of six parameters for \solc{} and ten parameters for \sole{}, which define the characteristics of the circles or ellipses, respectively.

The first step is to design the optimization objective. One option is to minimize the median error of the training micro-benchmarks. However, just minimizing the median might create a scenario as shown in Fig.~\ref{fig:defbin}, where runs with a high output probability of state $\ket{0}$ have a much lower error rate than runs with a low output probability. Therefore, to avoid this scenario the spread of the error among the runs (i.e., spread is defined as the 75$^{th}$ percentile error $-$ the 25$^{th}$ percentile error) is also minimized. However, if just the spread is chosen as the objective, then the median error might not be optimized (for example, spread can be minimized by choosing parameters such that all runs have an equally high error). \textit{Therefore, to create a balance between the median and the spread, the objective that \sol{} if to minimize the median error + the spread of the error} (we evaluate all three options in Sec.~\ref{sec:eval}).

Next, Algorithm~\ref{alg:discsa} shows how \sol{}'s simulated annealing engine iteratively steers toward a parameter configuration which minimizes the objective. Initially, the annealing environment has a high temperature, which means that it has a high likelihood of exploring different configuration neighborhoods. As the algorithm progresses, the temperature reduces and the algorithm narrows in on a near-optimal configuration neighborhood. At each iteration, a random configuration in the neighborhood of the current best configuration is sampled. A neighborhood is defined as all points within one unit for all of the parameters (six for \solc{} and ten for \sole{}). The objective of this sampled configuration is calculated for the training dataset. Depending on the current energy and temperature (as defined in Algorithm~\ref{alg:discsa}), it is determined if the best configuration should be updated to the newly sampled configuration. The best parameter configuration is obtained at the end of the execution, which can be used to classify the output of all succeeding quantum algorithm executions until the next calibration.

\textbf{Scaling to multiple qubits.} All qubits on a quantum computer need to be calibrated and separate classifiers need to be developed for all of them. The classification time overhead of \sol{} does not increase with the number of qubits as every step of \sol{} can run in parallel for all the qubits. This includes generating and executing the micro-benchmarks and independently running the simulated annealing engine concurrently. 
\section{\sol{} Evaluation}
\label{sec:eval}

\subsection{Evaluation Methodology}

We use IBM's Armonk quantum computing machine (specifications are provided in Table~\ref{tab:armonk}) to conduct our experiments. While IBM has other machines available, Armonk is the only one which allows obtaining the pre-classification raw complex values (OpenPulse).

For demonstrating robustness, \sol{} is evaluated and compared against IBM's existing state-of-art classifier over multiple days. The classifier's performance is evaluated using a validation dataset. The training and validation dataset are collected using the same methodology: 100 randomized U3-based micro-benchmarks, each with a different output probability of state $\ket{0}$, 10 belonging to each probability bin in increments of 0.1. In particular, \sol{} is evaluated across ten days, with a classifier developed for each day (each calibration). The two variants of \sol{}, \solc{} (circle classifier) and \sole{} (ellipse classifier), are compared against the default Qiskit classifier, referred to as the baseline classifier. The metrics used include the median error (of the 100 validation micro-benchmarks across the 10 days), the 75$^{th}$ percentile error, and the error spread (the 75$^{th}$ percentile error $-$ the 25$^{th}$ percentile error).

\subsection{Evaluation Analysis}

\begin{table}[t]
\centering
\caption{Specifications of Armonk Quantum Computer.}
\vspace{-0.3cm}
\scalebox{0.8}{
\begin{tabular}{|c|c|} 
 \hline
 Online Date & 10-16-2019 \\
 \hline
 Number of Qubits & 1 \\ 
 \hline
 Drive Frequency & $\approx$4.97 GHz \\
 \hline
 $\pi$ Pulse Shape & Gaussian \\
 \hline
 $\pi$ Pulse Amplitude & $\approx$0.252 + 0.0$j$  \\
 \hline
 $\pi$ Pulse Duration & $\approx$0.60 $\mu$s \\
 \hline
 Readout Frequency & $\approx$6.99 GHz \\
 \hline
 Readout Pulse Shape & Square Gaussian \\
 \hline
 Readout Pulse Amplitude & $\approx$0.605 + 0.0$j$ \\
 \hline
 Readout Pulse Duration & $\approx$3.52 $\mu$s \\
 \hline
\end{tabular}}
\label{tab:armonk}
\vspace{-0.4cm}
\end{table}

\begin{figure}[t]
    \centering
    \includegraphics[scale=0.47]{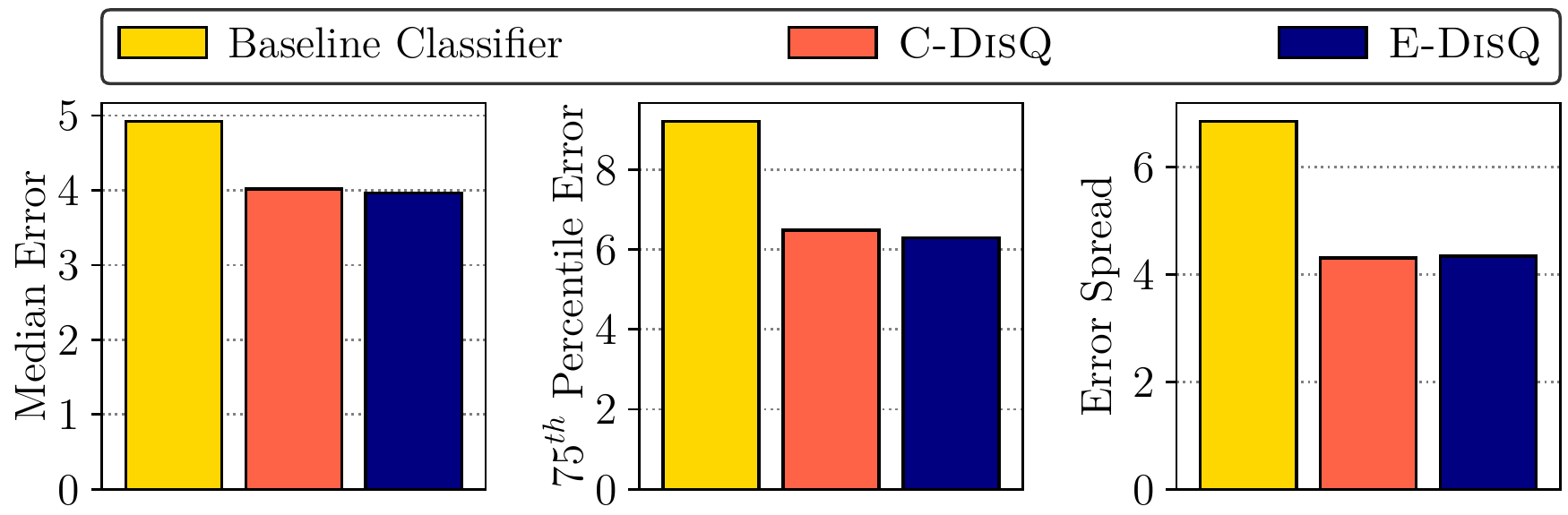}
    \vspace{0.1cm}
    \hrule
    \vspace{-0.3cm}
    \caption{Both variants of \sol{} achieve a lower median error, 75$^{th}$ percentile error, and error spread than the baseline classifier.}
    \vspace{-0.4cm}
    \label{fig:overal}
\end{figure}

\noindent\textbf{\sol{} achieves a lower median error, 75$^{th}$ percentile error, and error spread than the baseline classifier.} Fig.~\ref{fig:overal} shows that both \solc{} and \sole{} reduce the median error of the single-gate runs from 5\% to 4\%. Such a reduction can significantly improve the performance of quantum algorithms and programs with multiple gates. For instance, if five 5\%-error gates are applied, the overall error rate would be $1-(1-0.05)^5 = 23\%$. In comparison, if five 4\%-error gates are applied, the overall error rate would be $1-(1-0.04)^5 = 18\%$. Furthermore, the 75$^{th}$ percentile error is reduced from 9.5\% to 6.5\%. With the baseline classifier, 25\% of runs have an error rate of more than 9.5\%, while with \sol{}, 75\% of runs have an error rate of less than 6.5\% and 50\% of runs have an error of less than 4\%. Note that \sole{} performs slightly better than \solc{} for the median error as well as the 75$^{th}$ percentile error because it enables a higher level of precision over the classification region.

Next, the error spread drops from 7\% to $\approx$4.5\% with \solc{} and \sole{} -- a reduction of 36\%. This indicates that not only does \sol{} reduce the median error, but it also reduces the difference in the error rates observed by different runs, allowing for stable results. To follow up on this, next, we study how \sol{} reduces the difference in the error rates of runs with different output probabilities.

\begin{figure}
    \centering
    \includegraphics[scale=0.47]{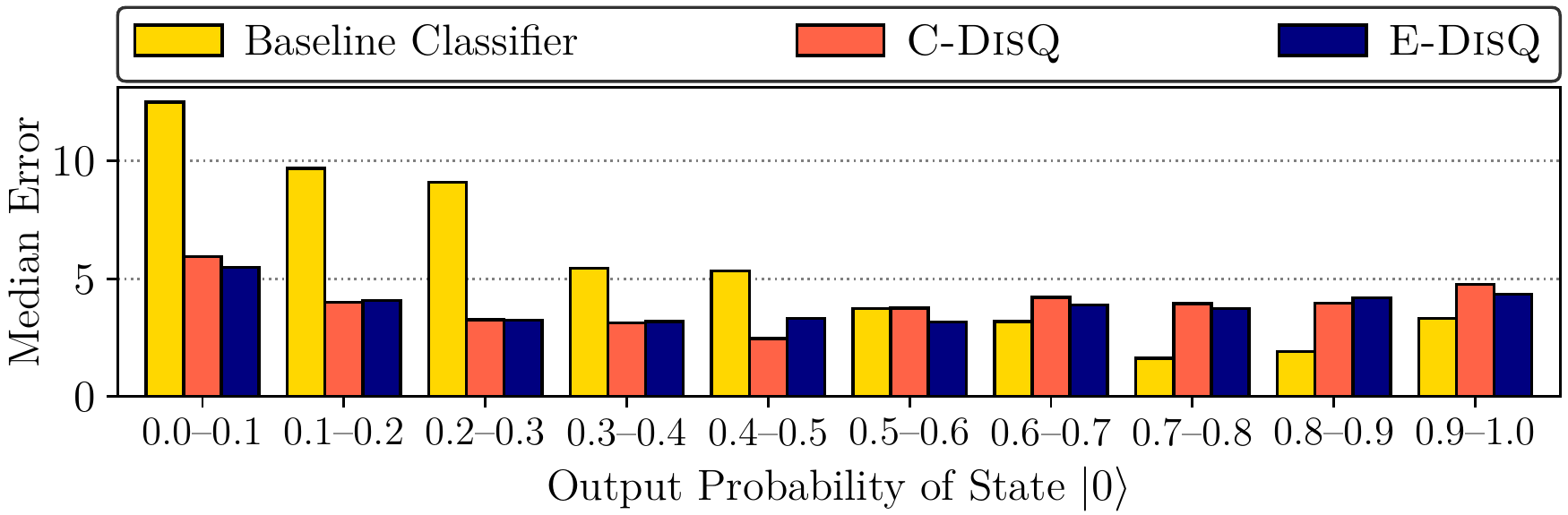}
    \vspace{0.1cm}
    \hrule
    \vspace{-0.3cm}
    \caption{\solc{} and \sole{} achieve a more equal error rate across runs with output probabilities of state $\ket{0}$ from 0 to 1.}
    \vspace{-0.6cm}
    \label{fig:medbin}
\end{figure}

\noindent\textbf{\newline \sol{} achieves a better equalization of the error rate compared to the baseline classifier.} Fig.~\ref{fig:medbin} shows the median errors for runs with different output probability of state $\ket{0}$ for the baseline classifier, \solc{}, and \sole{}. Both \solc{} and \sole{} have similar performance, achieve a more equal distribution of error for all output probabilities, especially reducing the error rate of runs with output probability of state $\ket{0}$ less than 0.5. For instance, the median error rate of the 0-0.1 bin is reduced from more than 12\% to less than 6\%. This is a reduction of more than 50\% in the error rate which has a significant impact on the stability of output of the runs with output probability of state $\ket{0}$ of less than 0.1. On the other hand, as a result of the distribution equalization, the error rates of runs with output probability of state $\ket{0}$ of more than 0.6 have increased. However, this increase is comparably smaller than the decrease in error rate for other runs. Overall, \sol{} achieves the goal of reducing the median error, while also equalizing the error rates across different outputs by also optimizing the error spread. Next, we look at the impact of optimizing different objective functions, including ones where the spread is not optimized.

\begin{figure}
    \centering
    \subfloat[][\solc{}: Circle Classifier]{\includegraphics[scale=0.47]{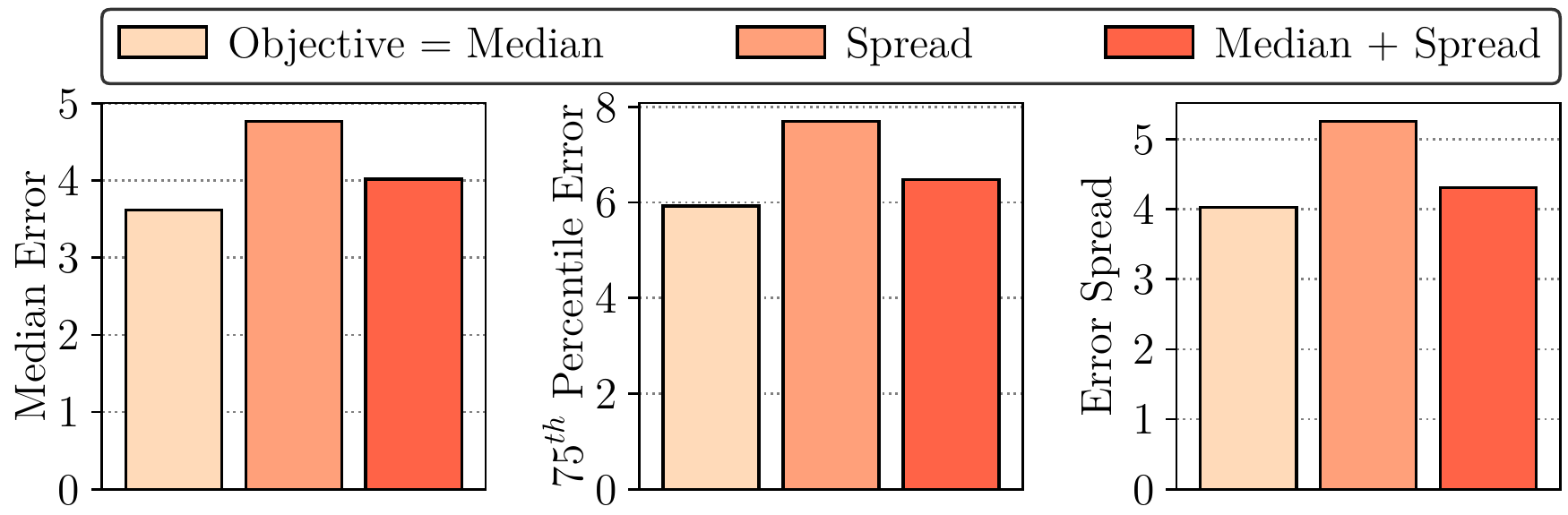}} \\
    \vspace{-0.2cm}
    \subfloat[][\sole{}: Ellipse Classifier]{\includegraphics[scale=0.47]{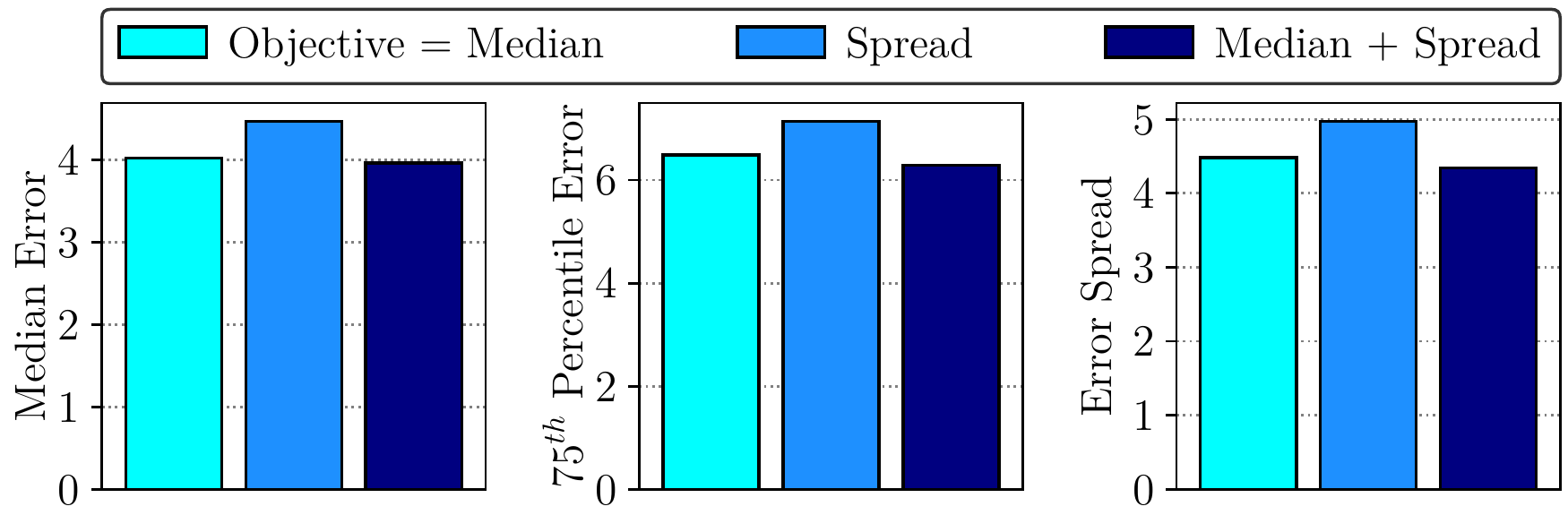}}
    \vspace{0.1cm}
    \hrule
    \vspace{-0.3cm}
    \caption{Optimizing just the median achieves similar median error and error spread as optimizing the median + the spread. Setting the objective to just the error spread performs worse.}
    \vspace{-0.4cm}
    \label{fig:mdnspd}
\end{figure}

\begin{figure}
    \centering
    \subfloat[][Objective = Median]{\includegraphics[scale=0.47]{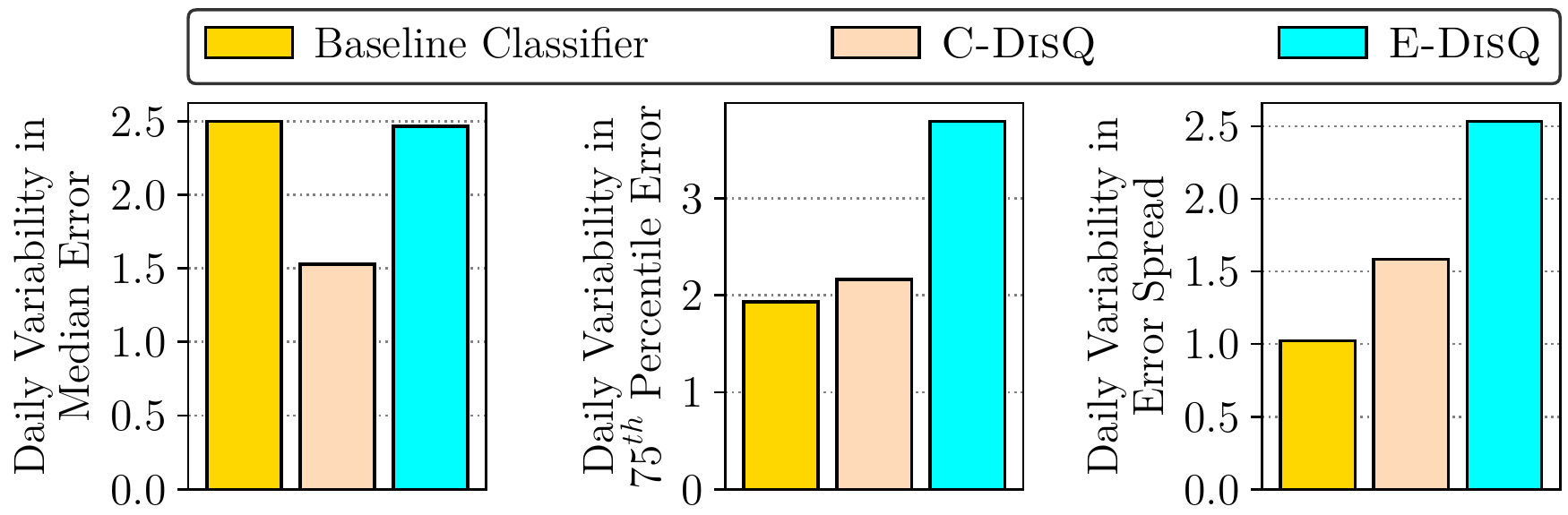}} \\
    \vspace{-0.2cm}
    \subfloat[][Objective = Median + Spread]{\includegraphics[scale=0.47]{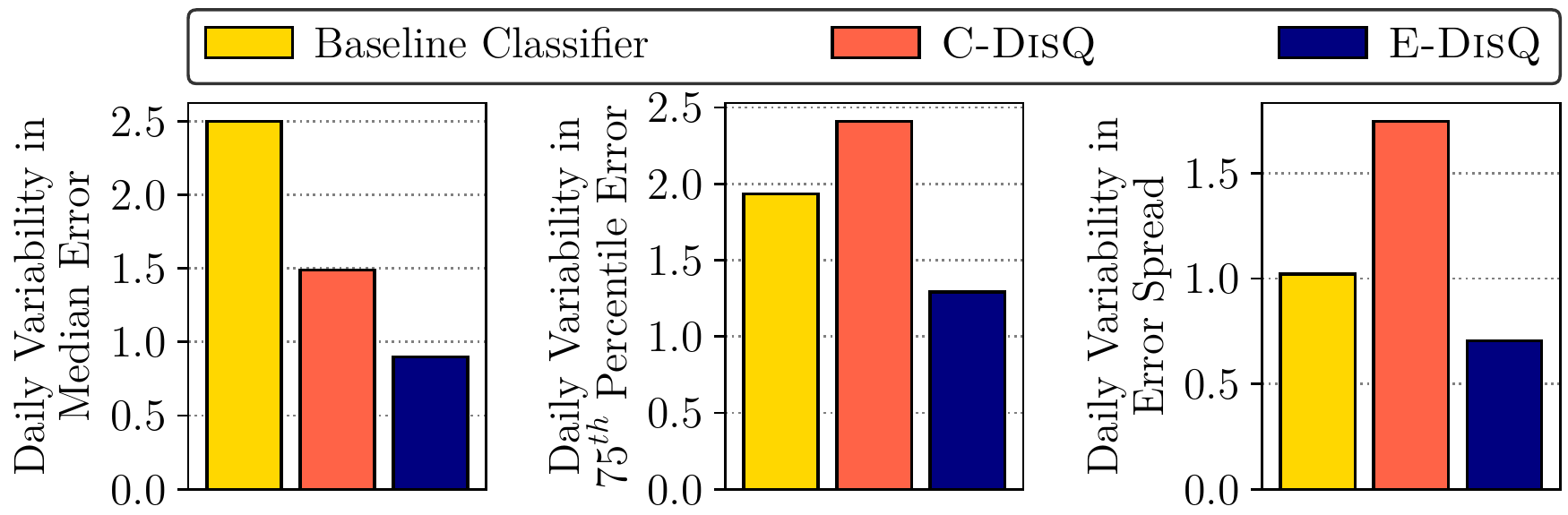}}
    \vspace{0.1cm}
    \hrule
    \vspace{-0.3cm}
    \caption{The day-to-day variability is lower with \sol{} when the objective is to minimize the median + the spread. When just the median is minimized, day-to-day variability is much higher.}
    \vspace{-0.4cm}
    \label{fig:dailys}
\end{figure}

\noindent\textbf{\newline Optimizing the median achieves similar low median error and error spread as optimizing the median + the spread. However, when the objective function is set to the median, day-to-day variability is much worse.} Fig.~\ref{fig:mdnspd}(a) and (b) show the median error, the 75$^{th}$ percentile error, and the error spread for choosing different objectives for the simulated annealing engine to optimize \solc{} and \sole{}, respectively. Evidently, in both cases, optimizing just the spread gives worse performance than optimizing the other two metrics. In fact, even the error spread is higher by over 1\% for \solc{} and 0.5\% for \sole{} when the error spread is optimized. The reason for this is that because optimizing the error spread does not focus on optimizing the median error, the configuration neighborhoods where the median is optimized are not explored. But those neighborhoods have the potential to reduce the spread just by reducing the median, as we observe in the case when the median is optimized. Optimizing the median gives similar results as optimizing the median + the spread, with the latter performing slightly worse for \solc{} and better for \sole{}.

However, this does not indicate that it is advisable to optimize just the median. Fig.~\ref{fig:dailys}(a) and (b) show the daily variability in error metrics when just the median is optimized and when the median + the spread is optimized, respectively. The daily variability is calculated as the spread of a given metric across the 10 days. For example, if the median error is considered, the median error is calculated for the validation dataset for each of the 10 days, and then, the spread of those 10 samples is indicated in the figure. The baseline case has a daily variability of 2\% error even in the case of the median error, which is a high variability. Interestingly, using the \sole{} classifier with the median + the spread as the objective reduces that variability to less than 1\%, which is a 50\% reduction in error variability. This makes the error rates more stable from one day to another, just improving the reproducibility of the results. However, using the \sole{} classifier with just the median as the objective has similar daily variability of median error as the baseline case. In fact, it has higher daily variability for the 75$^{th}$ percentile error and the error spread. This shows that while optimizing just the median error has good overall results, its daily variability is very high, making the results less reliable. On the other hand, \sol{} with the median + the spread as the objective, can reduce the daily variability and improve the stability considerably.

\begin{figure}
    \centering
    \subfloat[][\solc{}: Circle Classifier]{\includegraphics[scale=0.47]{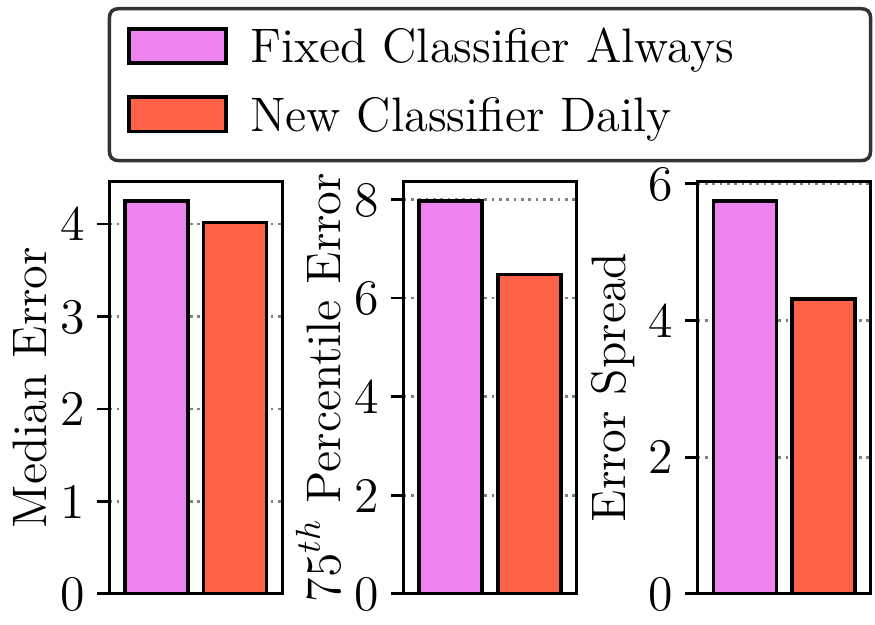}}
    \hfill
    \subfloat[][\sole{}: Ellipse Classifier]{\includegraphics[scale=0.47]{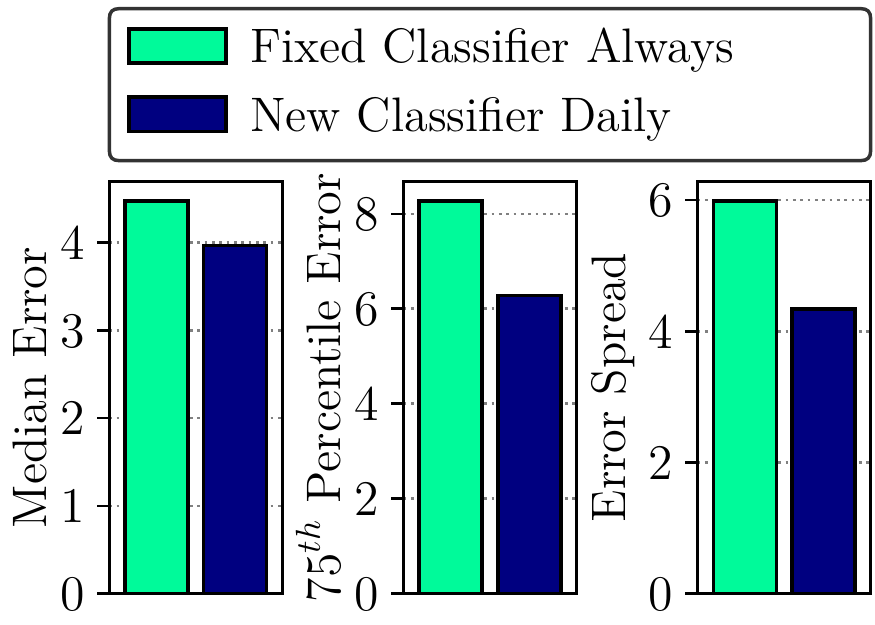}}
    \vspace{0.1cm}
    \hrule
    \vspace{-0.3cm}
    \caption{Developing a new classifier every calibration achieves lower error rate than using the same classifier across all calibrations.}
    \vspace{-0.4cm}
    \label{fig:bstnew}
\end{figure}

\begin{figure}
    \centering
    \includegraphics[scale=0.54]{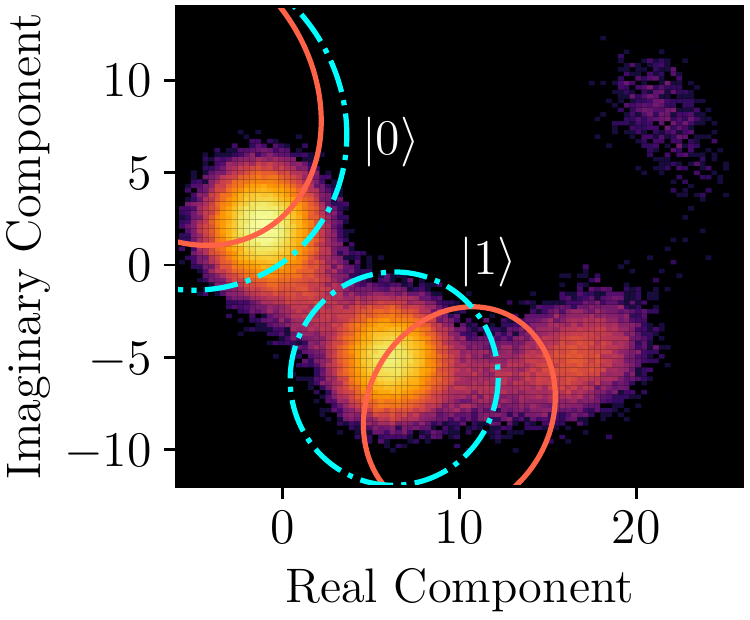}
    \hfill
    \includegraphics[scale=0.54]{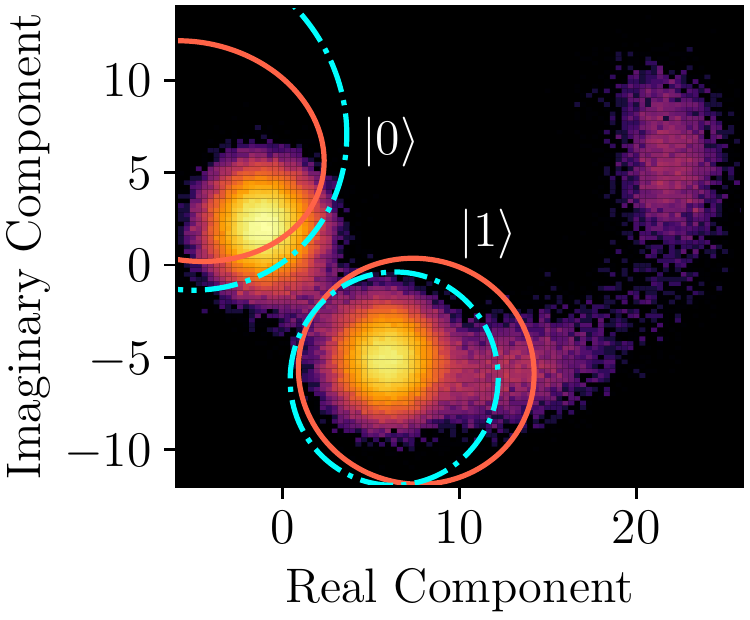}
    \vspace{0.1cm}
    \hrule
    \vspace{-0.3cm}
    \caption{Difference in classifiers when a fixed classifier is used always (broken blue lines) vs. when a new classifier is developed every calibration (red solid lines).}
    \vspace{-0.4cm}
    \label{fig:bsthst}
\end{figure}

\noindent\textbf{\newline Developing a new classifier every calibration achieves lower error rate and spread than using a fixed classifier across all calibrations, for both \solc{} and \sole{}.} One might question the significance of developing a new classifier every calibration, after all, the output of the same qubit is being classified. However, Fig.~\ref{fig:bstnew} shows the importance of developing a new classifier after each calibration. For both \solc{} and \sole{}, using a fixed classifier (the one which performs the best among all the ones generated over the 10 days) for all 10 calibrations performs worse than generating a new classifier after every calibration. For example, for \solc{} using a fixed classifier has an error spread of 6\%, while using a new classifier daily has an error spread of 4.3\%.

The reason for this is that due to the instability of the qubits, the measurement points generated daily might have different densities, requiring different classifiers as shown in Fig.~\ref{fig:bsthst}. Here, the broken blue line shows the fixed best classifier across the 10 days, while the red lines indicate the classifiers which were the best for the two days shown. Evidently, the regions considered by the fixed classifier are quite different than the regions considered by the day-specific classifiers, even though they have some intersecting portions. However, the disjoint portions are the ones which actually contribute to the higher error rate of the fixed classifier. 

\begin{figure}
    \centering
    \includegraphics[scale=0.47]{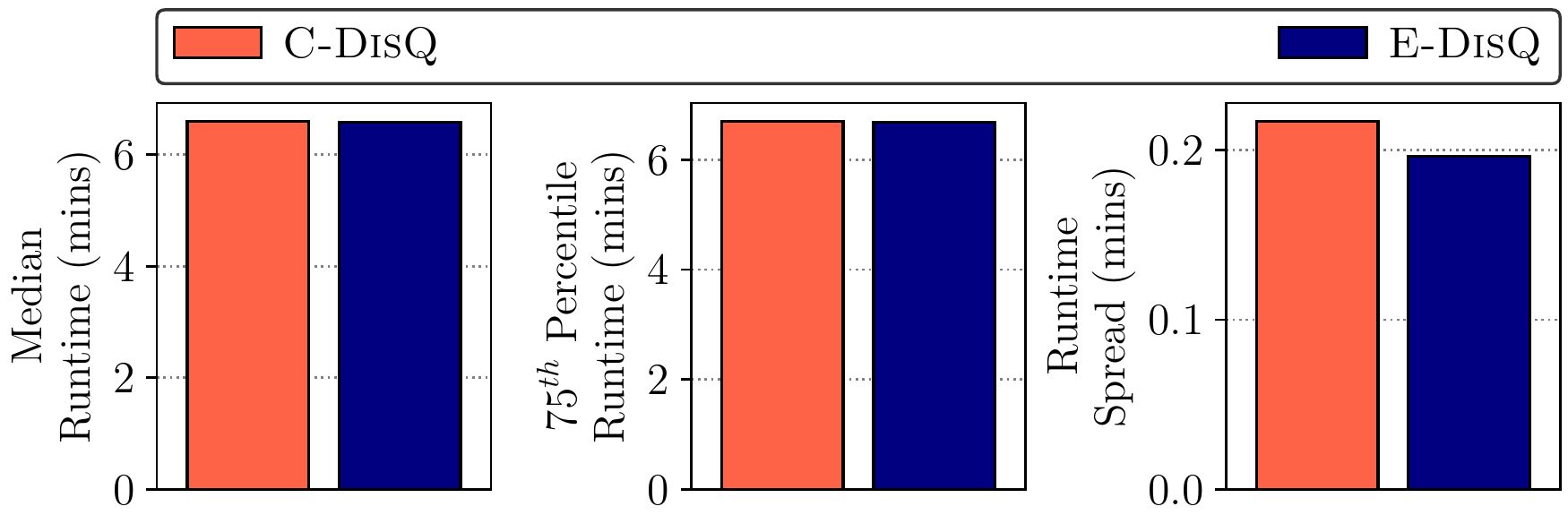}
    \vspace{0.1cm}
    \hrule
    \vspace{-0.3cm}
    \caption{Both \solc{} and \sole{} classifiers have similar low runtimes which are very stable across different executions.}
    \vspace{-0.4cm}
    \label{fig:runtme}
\end{figure}

\noindent\textbf{\newline Lastly, both \solc{} and \sole{} classifiers have similar low runtimes for executing the simulated annealing engine.} The overhead of running the micro-benchmarks on the quantum computer is very small as it can execute the entire batch of 100 runs, each with 1024 trials (100k trials in total),  within 2 minutes, which is comparable to the baseline case as most of the time is consumed in setting up and initializing the qubits. Once, the training dataset is generated, then the simulated annealing engine needs to be executed. Fig.~\ref{fig:runtme} shows that the execution time of \sol{}'s simulated annealing engine is less than 7 minutes on our local 4.20 GHz Intel Core i7-7700K machine for both \solc{} and \sole{} and the spread of these runtimes is $\approx$0.2 minutes across all the executions. This overhead is not on the critical path as the quantum computer can resume regular computation after the 2 minutes to run the micro-benchmarks. Note that we found that executing more than 100 micro-benchmarks (e.g., 1000) does not reduce the error rates, but does incur a higher overhead of performing the optimization.

Even though \sole{} optimizes 10 parameters, it has an equivalent runtime to \solc{}, which optimizes 6 parameters. Thus, overall, \sole{} not only has lower error rates, but also has comparable runtime to \solc{}, demonstrating that it is a more suitable option.
\section{Related Work}
\label{sec:relat}
Previous works have proposed qubit Quantum Error Correction (QEC) codes, which have an overhead of more than 10 physical qubits per logical qubit~\cite{sun2019experimental,layden2019ancilla,terhal2019scalable,burnett2019decoherence}. These methods are thus untenable for current NISQ devices, which require a low-overhead (in terms of the number of ancillary qubits required) solution to reduce the error so that they can execute quantum algorithms.

On the other hand, a large amount of focus has been dedicated toward optimizing the execution of quantum algorithms. This includes developing frameworks and compilers to optimize the mapping of a quantum algorithm to a quantum computer such as IBM's Qiskit compiler and Google's Cirq framework~\cite{aleksandrowiczqiskit,ahmed2018q,hancockcirq,computing2019pyquil,killoran2019strawberry}. In conjunction with these industry-led efforts, academic research has also focused on reducing the error rates of quantum algorithms by proposing debugging methods, simulation strategies, and noise-and-error-aware algorithm mapping approaches ~\cite{patel2020ureqa,li2020towards,bhattacharjee2019muqut,murali2020software,ash2019qure,das2019case,gokhale2019partial,huang2019statistical,li2019tackling,murali2019noise,murphy2019controlling,shi2019optimized,smith2019quantum,tannu2019mitigating,tannu2019not,smith2019quantum,butko2019understanding,wille2019mapping,zulehner2018efficient,zulehner2019compiling,mavadia2017prediction}. For example, Murali et al.~\cite{murali2020software} propose a method to reduce error rate by performing cross-talk-aware algorithm mapping to the qubits.

While these works focus on the higher-level problems of an algorithm's execution stack, IBM's OpenPulse, which is the framework to apply the pulses for quantum gates, has also been previously leveraged to solve problems including optimizing compilation~\cite{aleksandrowiczqiskit,dumitrescu2020benchmarking,capelluto2020openpulse,gokhale2020optimized}. However,  none of the above works have optimized the state classification problem by using raw output data of applying pulses, and thus, these works can be used in a compatible manner with \sol{} to minimize the error rates of NISQ devices.

\section{Conclusion}
\label{sec:concl}

This paper presented \sol{}, the first work to optimize state classifiers which differentiate between different quantum states. \sol{} demonstrates that the output error and its variability on NISQ devices can be reduced just by optimizing the classification methodology, without any hardware or compiler modifications. \sol{} is available at: \texttt{http://github.com/GoodwillComputingLab/DISQ}.

\small{\noindent\textbf{\\Acknowledgment.} We are thankful to anonymous reviewers for the constructive feedback and Northeastern University for supporting this work.}

\bibliographystyle{ACM-Reference-Format}
\bibliography{main}


\begin{thebibliography}{43}


\ifx \showCODEN    \undefined \def \showCODEN     #1{\unskip}     \fi
\ifx \showDOI      \undefined \def \showDOI       #1{#1}\fi
\ifx \showISBNx    \undefined \def \showISBNx     #1{\unskip}     \fi
\ifx \showISBNxiii \undefined \def \showISBNxiii  #1{\unskip}     \fi
\ifx \showISSN     \undefined \def \showISSN      #1{\unskip}     \fi
\ifx \showLCCN     \undefined \def \showLCCN      #1{\unskip}     \fi
\ifx \shownote     \undefined \def \shownote      #1{#1}          \fi
\ifx \showarticletitle \undefined \def \showarticletitle #1{#1}   \fi
\ifx \showURL      \undefined \def \showURL       {\relax}        \fi
\providecommand\bibfield[2]{#2}
\providecommand\bibinfo[2]{#2}
\providecommand\natexlab[1]{#1}
\providecommand\showeprint[2][]{arXiv:#2}

\bibitem[\protect\citeauthoryear{Ahmed}{Ahmed}{2018}]%
        {ahmed2018q}
\bibfield{author}{\bibinfo{person}{Talha Ahmed}.}
  \bibinfo{year}{2018}\natexlab{}.
\newblock \emph{\bibinfo{title}{Q\#: A Quantum Programming Language by
  Microsoft}}.
\newblock \bibinfo{thesistype}{Ph.D. Dissertation}. \bibinfo{school}{Imperial
  College London}.
\newblock


\bibitem[\protect\citeauthoryear{Aleksandrowicz, Alexander, Barkoutsos, Bello,
  Ben-Haim, Bucher, et~al\mbox{.}}{Aleksandrowicz et~al\mbox{.}}{[n.d.]}]%
        {aleksandrowiczqiskit}
\bibfield{author}{\bibinfo{person}{G Aleksandrowicz}, \bibinfo{person}{T
  Alexander}, \bibinfo{person}{P Barkoutsos}, \bibinfo{person}{L Bello},
  \bibinfo{person}{Y Ben-Haim}, \bibinfo{person}{D Bucher}, {et~al\mbox{.}}}
  \bibinfo{year}{[n.d.]}\natexlab{}.
\newblock \bibinfo{title}{{Qiskit: An Open-source Framework for Quantum
  Computing.(2019)}}.
\newblock
\newblock


\bibitem[\protect\citeauthoryear{Arute, Arya, Babbush, Bacon, Bardin, Barends,
  Biswas, Boixo, Brandao, Buell, et~al\mbox{.}}{Arute et~al\mbox{.}}{2019}]%
        {arute2019quantum}
\bibfield{author}{\bibinfo{person}{Frank Arute}, \bibinfo{person}{Kunal Arya},
  \bibinfo{person}{Ryan Babbush}, \bibinfo{person}{Dave Bacon},
  \bibinfo{person}{Joseph~C Bardin}, \bibinfo{person}{Rami Barends},
  \bibinfo{person}{Rupak Biswas}, \bibinfo{person}{Sergio Boixo},
  \bibinfo{person}{Fernando~GSL Brandao}, \bibinfo{person}{David~A Buell},
  {et~al\mbox{.}}} \bibinfo{year}{2019}\natexlab{}.
\newblock \showarticletitle{{Quantum Supremacy using a Programmable
  Superconducting Processor}}.
\newblock \bibinfo{journal}{\emph{Nature}} \bibinfo{volume}{574},
  \bibinfo{number}{7779} (\bibinfo{year}{2019}), \bibinfo{pages}{505--510}.
\newblock


\bibitem[\protect\citeauthoryear{Ash-Saki, Alam, and Ghosh}{Ash-Saki
  et~al\mbox{.}}{2019}]%
        {ash2019qure}
\bibfield{author}{\bibinfo{person}{Abdullah Ash-Saki},
  \bibinfo{person}{Mahabubul Alam}, {and} \bibinfo{person}{Swaroop Ghosh}.}
  \bibinfo{year}{2019}\natexlab{}.
\newblock \showarticletitle{{QURE: Qubit Re-allocation in Noisy
  Intermediate-Scale Quantum Computers}}. In
  \bibinfo{booktitle}{\emph{Proceedings of the 56th Annual Design Automation
  Conference 2019}}. ACM, \bibinfo{pages}{141}.
\newblock


\bibitem[\protect\citeauthoryear{Bhattacharjee, Saki, Alam, Chattopadhyay, and
  Ghosh}{Bhattacharjee et~al\mbox{.}}{2019}]%
        {bhattacharjee2019muqut}
\bibfield{author}{\bibinfo{person}{Debjyoti Bhattacharjee},
  \bibinfo{person}{Abdullah~Ash Saki}, \bibinfo{person}{Mahabubul Alam},
  \bibinfo{person}{Anupam Chattopadhyay}, {and} \bibinfo{person}{Swaroop
  Ghosh}.} \bibinfo{year}{2019}\natexlab{}.
\newblock \showarticletitle{{MUQUT: Multi-Constraint Quantum Circuit Mapping on
  NISQ Computers}}. In \bibinfo{booktitle}{\emph{38th IEEE/ACM International
  Conference on Computer-Aided Design, ICCAD 2019}}. Institute of Electrical
  and Electronics Engineers Inc., \bibinfo{pages}{8942132}.
\newblock


\bibitem[\protect\citeauthoryear{Bravyi, Smith, and Smolin}{Bravyi
  et~al\mbox{.}}{2016}]%
        {bravyi2016trading}
\bibfield{author}{\bibinfo{person}{Sergey Bravyi}, \bibinfo{person}{Graeme
  Smith}, {and} \bibinfo{person}{John~A Smolin}.}
  \bibinfo{year}{2016}\natexlab{}.
\newblock \showarticletitle{{Trading Classical and Quantum Computational
  Resources}}.
\newblock \bibinfo{journal}{\emph{Physical Review X}} \bibinfo{volume}{6},
  \bibinfo{number}{2} (\bibinfo{year}{2016}), \bibinfo{pages}{021043}.
\newblock


\bibitem[\protect\citeauthoryear{Burnett, Bengtsson, Scigliuzzo, Niepce, Kudra,
  Delsing, and Bylander}{Burnett et~al\mbox{.}}{2019}]%
        {burnett2019decoherence}
\bibfield{author}{\bibinfo{person}{Jonathan~J Burnett},
  \bibinfo{person}{Andreas Bengtsson}, \bibinfo{person}{Marco Scigliuzzo},
  \bibinfo{person}{David Niepce}, \bibinfo{person}{Marina Kudra},
  \bibinfo{person}{Per Delsing}, {and} \bibinfo{person}{Jonas Bylander}.}
  \bibinfo{year}{2019}\natexlab{}.
\newblock \showarticletitle{{Decoherence Benchmarking of Superconducting
  Qubits}}.
\newblock \bibinfo{journal}{\emph{npj Quantum Information}}
  \bibinfo{volume}{5}, \bibinfo{number}{1} (\bibinfo{year}{2019}).
\newblock


\bibitem[\protect\citeauthoryear{Butko, Michelogiannakis, Williams, Iancu,
  Donofrio, Shalf, Carter, and Siddiqi}{Butko et~al\mbox{.}}{2019}]%
        {butko2019understanding}
\bibfield{author}{\bibinfo{person}{Anastasiia Butko}, \bibinfo{person}{George
  Michelogiannakis}, \bibinfo{person}{Samuel Williams}, \bibinfo{person}{Costin
  Iancu}, \bibinfo{person}{David Donofrio}, \bibinfo{person}{John Shalf},
  \bibinfo{person}{Jonathan Carter}, {and} \bibinfo{person}{Irfan Siddiqi}.}
  \bibinfo{year}{2019}\natexlab{}.
\newblock \showarticletitle{Understanding Quantum Control Processor
  Capabilities and Limitations through Circuit Characterization}.
\newblock \bibinfo{journal}{\emph{arXiv preprint arXiv:1909.11719}}
  (\bibinfo{year}{2019}).
\newblock


\bibitem[\protect\citeauthoryear{Capelluto and Alexander}{Capelluto and
  Alexander}{2020}]%
        {capelluto2020openpulse}
\bibfield{author}{\bibinfo{person}{Lauren Capelluto} {and}
  \bibinfo{person}{Thomas Alexander}.} \bibinfo{year}{2020}\natexlab{}.
\newblock \showarticletitle{{OpenPulse: Software for Experimental Physicists in
  Quantum Computing}}.
\newblock \bibinfo{journal}{\emph{Bulletin of the American Physical Society}}
  (\bibinfo{year}{2020}).
\newblock


\bibitem[\protect\citeauthoryear{Cirac and Zoller}{Cirac and Zoller}{2012}]%
        {cirac2012goals}
\bibfield{author}{\bibinfo{person}{J~Ignacio Cirac} {and}
  \bibinfo{person}{Peter Zoller}.} \bibinfo{year}{2012}\natexlab{}.
\newblock \showarticletitle{{Goals and Opportunities in Quantum Simulation}}.
\newblock \bibinfo{journal}{\emph{Nature Physics}} \bibinfo{volume}{8},
  \bibinfo{number}{4} (\bibinfo{year}{2012}), \bibinfo{pages}{264}.
\newblock


\bibitem[\protect\citeauthoryear{Computing}{Computing}{2019}]%
        {computing2019pyquil}
\bibfield{author}{\bibinfo{person}{Rigetti Computing}.}
  \bibinfo{year}{2019}\natexlab{}.
\newblock \showarticletitle{Pyquil documentation}.
\newblock \bibinfo{journal}{\emph{URL http://pyquil. readthedocs.
  io/en/latest}} (\bibinfo{year}{2019}).
\newblock


\bibitem[\protect\citeauthoryear{Das, Tannu, Nair, and Qureshi}{Das
  et~al\mbox{.}}{2019}]%
        {das2019case}
\bibfield{author}{\bibinfo{person}{Poulami Das}, \bibinfo{person}{Swamit~S
  Tannu}, \bibinfo{person}{Prashant~J Nair}, {and} \bibinfo{person}{Moinuddin
  Qureshi}.} \bibinfo{year}{2019}\natexlab{}.
\newblock \showarticletitle{{A Case for Multi-Programming Quantum Computers}}.
  In \bibinfo{booktitle}{\emph{Proceedings of the 52nd Annual IEEE/ACM
  International Symposium on Microarchitecture}}. ACM,
  \bibinfo{pages}{291--303}.
\newblock


\bibitem[\protect\citeauthoryear{Dumitrescu, Pooser, and Garmon}{Dumitrescu
  et~al\mbox{.}}{2020}]%
        {dumitrescu2020benchmarking}
\bibfield{author}{\bibinfo{person}{Eugen Dumitrescu}, \bibinfo{person}{Raphael
  Pooser}, {and} \bibinfo{person}{John Garmon}.}
  \bibinfo{year}{2020}\natexlab{}.
\newblock \showarticletitle{{Benchmarking Noise Extrapolation with OpenPulse}}.
\newblock \bibinfo{journal}{\emph{Bulletin of the American Physical Society}}
  (\bibinfo{year}{2020}).
\newblock


\bibitem[\protect\citeauthoryear{Gokhale, Ding, Propson, Winkler, Leung, Shi,
  Schuster, Hoffmann, and Chong}{Gokhale et~al\mbox{.}}{2019}]%
        {gokhale2019partial}
\bibfield{author}{\bibinfo{person}{Pranav Gokhale}, \bibinfo{person}{Yongshan
  Ding}, \bibinfo{person}{Thomas Propson}, \bibinfo{person}{Christopher
  Winkler}, \bibinfo{person}{Nelson Leung}, \bibinfo{person}{Yunong Shi},
  \bibinfo{person}{David~I Schuster}, \bibinfo{person}{Henry Hoffmann}, {and}
  \bibinfo{person}{Frederic~T Chong}.} \bibinfo{year}{2019}\natexlab{}.
\newblock \showarticletitle{{Partial Compilation of Variational Algorithms for
  Noisy Intermediate-Scale Quantum Machines}}. In
  \bibinfo{booktitle}{\emph{Proceedings of the 52nd Annual IEEE/ACM
  International Symposium on Microarchitecture}}. ACM,
  \bibinfo{pages}{266--278}.
\newblock


\bibitem[\protect\citeauthoryear{Gokhale, Javadi-Abhari, Earnest, Shi, and
  Chong}{Gokhale et~al\mbox{.}}{2020}]%
        {gokhale2020optimized}
\bibfield{author}{\bibinfo{person}{Pranav Gokhale}, \bibinfo{person}{Ali
  Javadi-Abhari}, \bibinfo{person}{Nathan Earnest}, \bibinfo{person}{Yunong
  Shi}, {and} \bibinfo{person}{Frederic~T Chong}.}
  \bibinfo{year}{2020}\natexlab{}.
\newblock \showarticletitle{{Optimized Quantum Compilation for Near-Term
  Algorithms with OpenPulse}}.
\newblock \bibinfo{journal}{\emph{arXiv preprint arXiv:2004.11205}}
  (\bibinfo{year}{2020}).
\newblock


\bibitem[\protect\citeauthoryear{Hancock, Garcia, Shedenhelm, Cowen, and
  Carey}{Hancock et~al\mbox{.}}{[n.d.]}]%
        {hancockcirq}
\bibfield{author}{\bibinfo{person}{Andrew Hancock}, \bibinfo{person}{Austin
  Garcia}, \bibinfo{person}{Jacob Shedenhelm}, \bibinfo{person}{Jordan Cowen},
  {and} \bibinfo{person}{Calista Carey}.} \bibinfo{year}{[n.d.]}\natexlab{}.
\newblock \showarticletitle{Cirq: A Python Framework for Creating, Editing, and
  Invoking Quantum Circuits}.
\newblock  (\bibinfo{year}{[n.\,d.]}).
\newblock


\bibitem[\protect\citeauthoryear{Hassan, Pakin, and Feng}{Hassan
  et~al\mbox{.}}{2019}]%
        {hassan2019c}
\bibfield{author}{\bibinfo{person}{Mohamed~W Hassan}, \bibinfo{person}{Scott
  Pakin}, {and} \bibinfo{person}{Wu-chun Feng}.}
  \bibinfo{year}{2019}\natexlab{}.
\newblock \showarticletitle{{C to D-Wave: A High-level C Compilation Framework
  for Quantum Annealers}}. In \bibinfo{booktitle}{\emph{2019 IEEE High
  Performance Extreme Computing Conference (HPEC)}}. IEEE,
  \bibinfo{pages}{1--8}.
\newblock


\bibitem[\protect\citeauthoryear{Huang and Martonosi}{Huang and
  Martonosi}{2019}]%
        {huang2019statistical}
\bibfield{author}{\bibinfo{person}{Yipeng Huang} {and}
  \bibinfo{person}{Margaret Martonosi}.} \bibinfo{year}{2019}\natexlab{}.
\newblock \showarticletitle{{Statistical Assertions for Validating Patterns and
  Finding Bugs in Quantum Programs}}. In \bibinfo{booktitle}{\emph{Proceedings
  of the 46th International Symposium on Computer Architecture}}. ACM,
  \bibinfo{pages}{541--553}.
\newblock


\bibitem[\protect\citeauthoryear{Killoran, Izaac, Quesada, Bergholm, Amy, and
  Weedbrook}{Killoran et~al\mbox{.}}{2019}]%
        {killoran2019strawberry}
\bibfield{author}{\bibinfo{person}{Nathan Killoran}, \bibinfo{person}{Josh
  Izaac}, \bibinfo{person}{Nicol{\'a}s Quesada}, \bibinfo{person}{Ville
  Bergholm}, \bibinfo{person}{Matthew Amy}, {and} \bibinfo{person}{Christian
  Weedbrook}.} \bibinfo{year}{2019}\natexlab{}.
\newblock \showarticletitle{Strawberry fields: A software platform for photonic
  quantum computing}.
\newblock \bibinfo{journal}{\emph{Quantum}}  \bibinfo{volume}{3}
  (\bibinfo{year}{2019}), \bibinfo{pages}{129}.
\newblock


\bibitem[\protect\citeauthoryear{Layden, Zhou, Cappellaro, and Jiang}{Layden
  et~al\mbox{.}}{2019}]%
        {layden2019ancilla}
\bibfield{author}{\bibinfo{person}{David Layden}, \bibinfo{person}{Sisi Zhou},
  \bibinfo{person}{Paola Cappellaro}, {and} \bibinfo{person}{Liang Jiang}.}
  \bibinfo{year}{2019}\natexlab{}.
\newblock \showarticletitle{{Ancilla-Free Quantum Error Correction Codes for
  Quantum Metrology}}.
\newblock \bibinfo{journal}{\emph{Physical review letters}}
  \bibinfo{volume}{122}, \bibinfo{number}{4} (\bibinfo{year}{2019}).
\newblock


\bibitem[\protect\citeauthoryear{Li, Ding, and Xie}{Li et~al\mbox{.}}{2019}]%
        {li2019tackling}
\bibfield{author}{\bibinfo{person}{Gushu Li}, \bibinfo{person}{Yufei Ding},
  {and} \bibinfo{person}{Yuan Xie}.} \bibinfo{year}{2019}\natexlab{}.
\newblock \showarticletitle{{Tackling the Qubit Mapping Problem for NISQ-Era
  Quantum Devices}}. In \bibinfo{booktitle}{\emph{Proceedings of the
  Twenty-Fourth International Conference on Architectural Support for
  Programming Languages and Operating Systems}}. ACM,
  \bibinfo{pages}{1001--1014}.
\newblock


\bibitem[\protect\citeauthoryear{Li, Ding, and Xie}{Li et~al\mbox{.}}{2020}]%
        {li2020towards}
\bibfield{author}{\bibinfo{person}{Gushu Li}, \bibinfo{person}{Yufei Ding},
  {and} \bibinfo{person}{Yuan Xie}.} \bibinfo{year}{2020}\natexlab{}.
\newblock \showarticletitle{Towards Efficient Superconducting Quantum Processor
  Architecture Design}. In \bibinfo{booktitle}{\emph{Proceedings of the
  Twenty-Fifth International Conference on Architectural Support for
  Programming Languages and Operating Systems}}. \bibinfo{pages}{1031--1045}.
\newblock


\bibitem[\protect\citeauthoryear{Liu, Byrd, and Zhou}{Liu
  et~al\mbox{.}}{2020}]%
        {liu2020quantum}
\bibfield{author}{\bibinfo{person}{Ji Liu}, \bibinfo{person}{Gregory~T Byrd},
  {and} \bibinfo{person}{Huiyang Zhou}.} \bibinfo{year}{2020}\natexlab{}.
\newblock \showarticletitle{Quantum Circuits for Dynamic Runtime Assertions in
  Quantum Computation}. In \bibinfo{booktitle}{\emph{Proceedings of the
  Twenty-Fifth International Conference on Architectural Support for
  Programming Languages and Operating Systems}}. \bibinfo{pages}{1017--1030}.
\newblock


\bibitem[\protect\citeauthoryear{Martonosi and Roetteler}{Martonosi and
  Roetteler}{2019}]%
        {martonosi2019next}
\bibfield{author}{\bibinfo{person}{Margaret Martonosi} {and}
  \bibinfo{person}{Martin Roetteler}.} \bibinfo{year}{2019}\natexlab{}.
\newblock \showarticletitle{{Next Steps in Quantum Computing: Computer
  Science's Role}}.
\newblock \bibinfo{journal}{\emph{arXiv preprint arXiv:1903.10541}}
  (\bibinfo{year}{2019}).
\newblock


\bibitem[\protect\citeauthoryear{Mavadia et~al\mbox{.}}{Mavadia
  et~al\mbox{.}}{2017}]%
        {mavadia2017prediction}
\bibfield{author}{\bibinfo{person}{Mavadia} {et~al\mbox{.}}}
  \bibinfo{year}{2017}\natexlab{}.
\newblock \showarticletitle{{Prediction and Real-Time Compensation of Qubit
  Decoherence via Machine Learning}}.
\newblock \bibinfo{journal}{\emph{Nature communications}}  \bibinfo{volume}{8}
  (\bibinfo{year}{2017}).
\newblock


\bibitem[\protect\citeauthoryear{McKay, Alexander, Bello, Biercuk, Bishop,
  Chen, Chow, C{\'o}rcoles, Egger, Filipp, et~al\mbox{.}}{McKay
  et~al\mbox{.}}{2018}]%
        {mckay2018qiskit}
\bibfield{author}{\bibinfo{person}{David~C McKay}, \bibinfo{person}{Thomas
  Alexander}, \bibinfo{person}{Luciano Bello}, \bibinfo{person}{Michael~J
  Biercuk}, \bibinfo{person}{Lev Bishop}, \bibinfo{person}{Jiayin Chen},
  \bibinfo{person}{Jerry~M Chow}, \bibinfo{person}{Antonio~D C{\'o}rcoles},
  \bibinfo{person}{Daniel Egger}, \bibinfo{person}{Stefan Filipp},
  {et~al\mbox{.}}} \bibinfo{year}{2018}\natexlab{}.
\newblock \showarticletitle{{Qiskit Backend Specifications for OpenQASM and
  OpenPulse Experiments}}.
\newblock \bibinfo{journal}{\emph{arXiv preprint arXiv:1809.03452}}
  (\bibinfo{year}{2018}).
\newblock


\bibitem[\protect\citeauthoryear{Murali, Baker, Javadi-Abhari, Chong, and
  Martonosi}{Murali et~al\mbox{.}}{2019}]%
        {murali2019noise}
\bibfield{author}{\bibinfo{person}{Prakash Murali}, \bibinfo{person}{Jonathan~M
  Baker}, \bibinfo{person}{Ali Javadi-Abhari}, \bibinfo{person}{Frederic~T
  Chong}, {and} \bibinfo{person}{Margaret Martonosi}.}
  \bibinfo{year}{2019}\natexlab{}.
\newblock \showarticletitle{{Noise-Adaptive Compiler Mappings for Noisy
  Intermediate-Scale Quantum Computers}}. In
  \bibinfo{booktitle}{\emph{Proceedings of the Twenty-Fourth International
  Conference on Architectural Support for Programming Languages and Operating
  Systems}}. ACM, \bibinfo{pages}{1015--1029}.
\newblock


\bibitem[\protect\citeauthoryear{Murali, McKay, Martonosi, and
  Javadi-Abhari}{Murali et~al\mbox{.}}{2020}]%
        {murali2020software}
\bibfield{author}{\bibinfo{person}{Prakash Murali}, \bibinfo{person}{David~C
  McKay}, \bibinfo{person}{Margaret Martonosi}, {and} \bibinfo{person}{Ali
  Javadi-Abhari}.} \bibinfo{year}{2020}\natexlab{}.
\newblock \showarticletitle{Software Mitigation of Crosstalk on Noisy
  Intermediate-Scale Quantum Computers}.
\newblock  (\bibinfo{year}{2020}), \bibinfo{pages}{1003--1016}.
\newblock


\bibitem[\protect\citeauthoryear{Murphy and Brown}{Murphy and Brown}{2019}]%
        {murphy2019controlling}
\bibfield{author}{\bibinfo{person}{Daniel~C Murphy} {and}
  \bibinfo{person}{Kenneth~R Brown}.} \bibinfo{year}{2019}\natexlab{}.
\newblock \showarticletitle{{Controlling Error Orientation to Improve Quantum
  Algorithm Success Rates}}.
\newblock \bibinfo{journal}{\emph{Physical Review A}} \bibinfo{volume}{99},
  \bibinfo{number}{3} (\bibinfo{year}{2019}), \bibinfo{pages}{032318}.
\newblock


\bibitem[\protect\citeauthoryear{Pakin}{Pakin}{2016}]%
        {pakin2016quantum}
\bibfield{author}{\bibinfo{person}{Scott Pakin}.}
  \bibinfo{year}{2016}\natexlab{}.
\newblock \showarticletitle{{A Quantum Macro Assembler}}. In
  \bibinfo{booktitle}{\emph{2016 IEEE High Performance Extreme Computing
  Conference (HPEC)}}. IEEE, \bibinfo{pages}{1--8}.
\newblock


\bibitem[\protect\citeauthoryear{Pakin and Reinhardt}{Pakin and
  Reinhardt}{2018}]%
        {pakin2018survey}
\bibfield{author}{\bibinfo{person}{Scott Pakin} {and} \bibinfo{person}{Steven~P
  Reinhardt}.} \bibinfo{year}{2018}\natexlab{}.
\newblock \showarticletitle{{A Survey of Programming Tools for D-Wave
  Quantum-Annealing Processors}}. In \bibinfo{booktitle}{\emph{International
  Conference on High Performance Computing}}. Springer,
  \bibinfo{pages}{103--122}.
\newblock


\bibitem[\protect\citeauthoryear{Patel, Li, Roy, and Tiwari}{Patel
  et~al\mbox{.}}{2020}]%
        {patel2020ureqa}
\bibfield{author}{\bibinfo{person}{Tirthak Patel}, \bibinfo{person}{Baolin Li},
  \bibinfo{person}{Rohan~Basu Roy}, {and} \bibinfo{person}{Devesh Tiwari}.}
  \bibinfo{year}{2020}\natexlab{}.
\newblock \showarticletitle{$\{$UREQA$\}$: Leveraging Operation-Aware Error
  Rates for Effective Quantum Circuit Mapping on NISQ-Era Quantum Computers}.
  In \bibinfo{booktitle}{\emph{2020 $\{$USENIX$\}$ Annual Technical Conference
  ($\{$USENIX$\}\{$ATC$\}$ 20)}}. \bibinfo{pages}{705--711}.
\newblock


\bibitem[\protect\citeauthoryear{Preskill}{Preskill}{2018}]%
        {preskill2018quantum}
\bibfield{author}{\bibinfo{person}{John Preskill}.}
  \bibinfo{year}{2018}\natexlab{}.
\newblock \showarticletitle{{Quantum Computing in the NISQ Era and Beyond}}.
\newblock \bibinfo{journal}{\emph{Quantum}}  \bibinfo{volume}{2}
  (\bibinfo{year}{2018}), \bibinfo{pages}{79}.
\newblock


\bibitem[\protect\citeauthoryear{Shi, Leung, Gokhale, Rossi, Schuster,
  Hoffmann, and Chong}{Shi et~al\mbox{.}}{2019}]%
        {shi2019optimized}
\bibfield{author}{\bibinfo{person}{Yunong Shi}, \bibinfo{person}{Nelson Leung},
  \bibinfo{person}{Pranav Gokhale}, \bibinfo{person}{Zane Rossi},
  \bibinfo{person}{David~I Schuster}, \bibinfo{person}{Henry Hoffmann}, {and}
  \bibinfo{person}{Frederic~T Chong}.} \bibinfo{year}{2019}\natexlab{}.
\newblock \showarticletitle{{Optimized Compilation of Aggregated Instructions
  for Realistic Quantum Computers}}. In \bibinfo{booktitle}{\emph{Proceedings
  of the Twenty-Fourth International Conference on Architectural Support for
  Programming Languages and Operating Systems}}. ACM,
  \bibinfo{pages}{1031--1044}.
\newblock


\bibitem[\protect\citeauthoryear{Smith and Thornton}{Smith and
  Thornton}{2019}]%
        {smith2019quantum}
\bibfield{author}{\bibinfo{person}{Kaitlin~N Smith} {and}
  \bibinfo{person}{Mitchell~A Thornton}.} \bibinfo{year}{2019}\natexlab{}.
\newblock \showarticletitle{{A Quantum Computational Compiler and Design Tool
  for Technology-Specific Targets}}. In \bibinfo{booktitle}{\emph{Proceedings
  of the 46th International Symposium on Computer Architecture}}. ACM,
  \bibinfo{pages}{579--588}.
\newblock


\bibitem[\protect\citeauthoryear{Sun, Hu, Ma, Cai, Mu, Xu, Weiting, Wu, Wang,
  Song, et~al\mbox{.}}{Sun et~al\mbox{.}}{2019}]%
        {sun2019experimental}
\bibfield{author}{\bibinfo{person}{Luyan Sun}, \bibinfo{person}{Ling Hu},
  \bibinfo{person}{Yuwei Ma}, \bibinfo{person}{Weizhou Cai},
  \bibinfo{person}{Xianghao Mu}, \bibinfo{person}{Yuan Xu},
  \bibinfo{person}{Wang Weiting}, \bibinfo{person}{Yukai Wu},
  \bibinfo{person}{Haiyan Wang}, \bibinfo{person}{Yipu Song}, {et~al\mbox{.}}}
  \bibinfo{year}{2019}\natexlab{}.
\newblock \showarticletitle{{Experimental Quantum Error Correction with
  Binomial Bosonic Codes}}. In \bibinfo{booktitle}{\emph{APS Meeting
  Abstracts}}.
\newblock


\bibitem[\protect\citeauthoryear{Tannu and Qureshi}{Tannu and Qureshi}{2019a}]%
        {tannu2019ensemble}
\bibfield{author}{\bibinfo{person}{Swamit~S Tannu} {and}
  \bibinfo{person}{Moinuddin Qureshi}.} \bibinfo{year}{2019}\natexlab{a}.
\newblock \showarticletitle{{Ensemble of Diverse Mappings: Improving
  Reliability of Quantum Computers by Orchestrating Dissimilar Mistakes}}. In
  \bibinfo{booktitle}{\emph{Proceedings of the 52nd Annual IEEE/ACM
  International Symposium on Microarchitecture}}. ACM,
  \bibinfo{pages}{253--265}.
\newblock


\bibitem[\protect\citeauthoryear{Tannu and Qureshi}{Tannu and Qureshi}{2019b}]%
        {tannu2019mitigating}
\bibfield{author}{\bibinfo{person}{Swamit~S Tannu} {and}
  \bibinfo{person}{Moinuddin~K Qureshi}.} \bibinfo{year}{2019}\natexlab{b}.
\newblock \showarticletitle{{Mitigating Measurement Errors in Quantum Computers
  by Exploiting State-Dependent Bias}}. In
  \bibinfo{booktitle}{\emph{Proceedings of the 52nd Annual IEEE/ACM
  International Symposium on Microarchitecture}}. ACM,
  \bibinfo{pages}{279--290}.
\newblock


\bibitem[\protect\citeauthoryear{Tannu and Qureshi}{Tannu and Qureshi}{2019c}]%
        {tannu2019not}
\bibfield{author}{\bibinfo{person}{Swamit~S Tannu} {and}
  \bibinfo{person}{Moinuddin~K Qureshi}.} \bibinfo{year}{2019}\natexlab{c}.
\newblock \showarticletitle{{Not All Aubits are Created Equal: A Case for
  Variability-Aware Policies for NISQ-Era Quantum Computers}}. In
  \bibinfo{booktitle}{\emph{Proceedings of the Twenty-Fourth International
  Conference on Architectural Support for Programming Languages and Operating
  Systems}}. ACM, \bibinfo{pages}{987--999}.
\newblock


\bibitem[\protect\citeauthoryear{Terhal, Pryadko, Weigand, Wang, Asasi, and
  Vuillot}{Terhal et~al\mbox{.}}{2019}]%
        {terhal2019scalable}
\bibfield{author}{\bibinfo{person}{Barbara Terhal}, \bibinfo{person}{Leonid
  Pryadko}, \bibinfo{person}{Daniel Weigand}, \bibinfo{person}{Yang Wang},
  \bibinfo{person}{Hamed Asasi}, {and} \bibinfo{person}{Christophe Vuillot}.}
  \bibinfo{year}{2019}\natexlab{}.
\newblock \showarticletitle{{Scalable Quantum Error Correction with the Bosonic
  GKP Code}}. In \bibinfo{booktitle}{\emph{APS Meeting Abstracts}}.
\newblock


\bibitem[\protect\citeauthoryear{Wille, Burgholzer, and Zulehner}{Wille
  et~al\mbox{.}}{2019}]%
        {wille2019mapping}
\bibfield{author}{\bibinfo{person}{Robert Wille}, \bibinfo{person}{Lukas
  Burgholzer}, {and} \bibinfo{person}{Alwin Zulehner}.}
  \bibinfo{year}{2019}\natexlab{}.
\newblock \showarticletitle{{Mapping Quantum Circuits to IBM QX Architectures
  Using the Minimal Number of SWAP and H Operations}}. In
  \bibinfo{booktitle}{\emph{Proceedings of the 56th Annual Design Automation
  Conference 2019}}. ACM, \bibinfo{pages}{142}.
\newblock


\bibitem[\protect\citeauthoryear{Zulehner, Paler, and Wille}{Zulehner
  et~al\mbox{.}}{2018}]%
        {zulehner2018efficient}
\bibfield{author}{\bibinfo{person}{Alwin Zulehner}, \bibinfo{person}{Alexandru
  Paler}, {and} \bibinfo{person}{Robert Wille}.}
  \bibinfo{year}{2018}\natexlab{}.
\newblock \showarticletitle{{An Efficient Methodology for Mapping Quantum
  Circuits to the IBM QX Architectures}}.
\newblock \bibinfo{journal}{\emph{IEEE Transactions on Computer-Aided Design of
  Integrated Circuits and Systems}} (\bibinfo{year}{2018}).
\newblock


\bibitem[\protect\citeauthoryear{Zulehner and Wille}{Zulehner and
  Wille}{2019}]%
        {zulehner2019compiling}
\bibfield{author}{\bibinfo{person}{Alwin Zulehner} {and}
  \bibinfo{person}{Robert Wille}.} \bibinfo{year}{2019}\natexlab{}.
\newblock \showarticletitle{{Compiling SU (4) Quantum Circuits to IBM QX
  Architectures}}. In \bibinfo{booktitle}{\emph{Proceedings of the 24th Asia
  and South Pacific Design Automation Conference}}. ACM,
  \bibinfo{pages}{185--190}.
\newblock


\end{thebibliography}

\end{document}